\documentclass[twocolumn,preprintnumbers,amsmath,amssymb,pre]{revtex4}

\usepackage{graphicx}
\usepackage{subfig}
\usepackage{dcolumn}
\usepackage{bm}
\usepackage{marvosym}

\usepackage{color}
\definecolor{light-gray}{gray}{0.5}
\definecolor{blue}{rgb}{0.0,0.0,1.0}
\definecolor{green}{rgb}{0.0,0.5,0.0}
\definecolor{red}{rgb}{1.0,0.0,0.0}
\definecolor{cyan}{rgb}{0.0,0.75,0.75}
\definecolor{magenta}{rgb}{0.75,0.0,0.75}
\definecolor{yellow}{rgb}{0.75,0.75,0.0}

\newcommand{\avg}[1]{\langle{#1}\rangle}
\newcommand{\sdot}{\cdot}
\newcommand{\ox}{\otimes}
\newcommand{\grad}{\bm \nabla}
\newcommand{\pd}{\partial}

\newcommand{\RE}{\mathcal{R}e}

\newcommand{\etal}{\emph{et al.}}

\newcommand{\lt}{\left}
\newcommand{\rt}{\right}

\newcommand{\dd}{\mathrm{d}}
\newcommand{\Lint}{\big|_{\mathcal L}}

\begin{document}
\title{The stagnation point von K\'arm\'an coefficient}
\author{V. Dallas}
\email{vassilios.dallas04@imperial.ac.uk}
\affiliation{Institute for Mathematical Sciences, Imperial College, London, SW7 2PG, UK \\ Department of Aeronautics, Imperial College, London, SW7 2AZ, UK}
\author{J. C. Vassilicos}
\email{j.c.vassilicos@imperial.ac.uk}
\affiliation{Institute for Mathematical Sciences, Imperial College, London, SW7 2PG, UK \\ Department of Aeronautics, Imperial College, London, SW7 2AZ, UK}
\author{G. F. Hewitt}
\affiliation{Department of Chemical Engineering and Chemical Technology, Imperial College, London, SW7 2AZ, UK}
\begin{abstract}
On the basis of various Direct Numerical Simulations (DNS) of turbulent channel flows the following picture is proposed. (\emph i) At a height $y$ from either wall, the Taylor microscale $\lambda$ is proportional to the average distance $\ell_s$ between stagnation points of the fluctuating velocity field, i.e. $\lambda(y) = B_1 \ell_s(y)$ with $B_1$ constant, for $\delta_\nu \ll y \lesssim \delta$, where the wall unit $\delta_\nu$ is defined as the ratio of kinematic viscosity $\nu$ to skin friction velocity $u_\tau$ and $\delta$ is the channel's half width. (\emph{ii}) The number density $n_s$ of stagnation points varies with height according to $n_s = \frac{C_s}{\delta_\nu^3}y_+^{-1}$ where $y_+ \equiv y/\delta_\nu$ and $C_s$ is constant in the range $\delta_\nu \ll y \lesssim \delta$. (\emph{iii}) In that same range, the kinetic energy dissipation rate per unit mass, $\epsilon$, equals $\frac{2}{3} E_+ \frac{u_\tau^3}{\kappa_s y}$ where $E_+$ is the total kinetic energy per unit mass normalised by $u_\tau^2$ and $\kappa_s = B_1^2 / C_s$ is the stagnation point von K\'arm\'an coefficient. (\emph{iv}) In the limit of exceedingly large Reynolds numbers $\RE_\tau \equiv \delta / \delta_\nu$, large enough for the Reynolds stress $-\avg{uv}$ to equal $u_\tau^2$ in the range $\delta_\nu \ll y \ll \delta$, and assuming that production of turbulent kinetic energy balances dissipation locally in that range and limit, the mean velocity $U_+$, normalised by $u_\tau$, obeys $\frac{\dd}{\dd y} U_+ \simeq \frac{2}{3} \frac{E_+}{\kappa_s y}$ in that same range. (\emph{v}) It follows that the von K\'arm\'an coefficient $\kappa$ is a meaningful and well-defined coefficient and the log-law holds in turbulent channel/pipe flows only if $E_+$ is independent of $y_+$ and $\RE_\tau$ in that range, in which case $\kappa \sim \kappa_s$.  (\emph{vi}) In support of $\frac{\dd}{\dd y} U_+ \simeq \frac{2}{3} \frac{E_+}{\kappa_s y}$, DNS data of turbulent channel flows which include the highest currently available values of $\RE_\tau$ are best fitted by $E_+ \simeq \frac{2}{3} B_4 y_+^{-2/15}$ and $\frac{\dd}{\dd y_+}U_+ \simeq \frac{B_4}{\kappa_s} y_+^{-1 - 2/15}$ with $B_4$ independent of $y$ in $\delta_\nu \ll y \ll \delta$ if the significant departure from $-\avg{uv} \simeq u_\tau^2$ at these $\RE_\tau$ values is taken into account. 
\end{abstract}
\maketitle
\section{\label{sec:intro}Introduction}
The mean flow profile of turbulent boundary layers is very widely taken to incorporate an extensive log-law region for a very broad range of turbulent wall-bounded flows. As a result, the log-law and its so-called von K\'arm\'an constant are a central part of most engineering turbulence models. However, renewed interest and new measurements over the past ten to fifteen years have led to debates (\emph i) on the form of the mean velocity profile (is it a log-law or a power-law with very weak power exponent?), (\emph{ii}) on its scalings with Reynolds number and (\emph{iii}) on its dependence on or independence of overall flow geometry (see the Theme Issue of Phil. Trans. R. Soc. Lond. A (2007), volume 365, on ``Scaling and structure in high Reynolds number wall bounded flows''). Fittings of new mean flow data with a log-law lead to a variety of values for the von K\'arm\'an constant $\kappa$: as low as $1/e$ for channel flows \cite{zanounetal03}, as high as 0.43 for pipe flows \cite{zagarolasmits98} and $\kappa \simeq 0.38$ for zero-pressure-gradient boundary layer flows \cite{nagibchauhan08}. Due to this non-universality the von K\'arm\'an constant is being renamed von K\'arm\'an coefficient \cite{nagibchauhan08}.

These seemingly small departures from the classically accepted \cite{pope00} $\kappa \simeq 0.41$ value matter because of the reliance that so many turbulence models have on the log-law. In a personal communication in early 2009, Phillip Spalart mentions that ``the recent fluctuations in preferred $\kappa$ values leads to changes of the order of 1\% in the drag predicted by CFD for an airplane with typical RANS models''. In terms of fuel economy sustained over years, this 1\% drag change is significant. 

All these are issues which in one way or another would have been solved if the turbulence closure problem had been solved. In this respect, the main goal of the present paper is not to find definitive answers to these specific issues, nor is it to find closure relations between different order statistics of flow quantities \cite{lvovetal08}. Instead, our main goal is to determine relations between mean flow quantities, such as the mean flow and the dissipation rate of kinetic energy, and the underlying flow field topography of the fluctuating velocities. Recent results obtained for homogeneous isotropic turbulence indicate that it is not impossible to find relations between bulk flow statistics and the underlying topography of the fluctuating velocity field. Mazellier and Vassilicos \cite{mv08} related the dissipation constant $C_{\epsilon}$ to the number of zero-crossings of velocity fluctuations and were able to take account the non-universality of $C_{\epsilon}$ by their formula. Goto and Vassilicos \cite{gv09} went one step further and related $C_{\epsilon}$ to the number of stagnation points of the velocity fluctuations. The stagnation points are the most basic aspect of field topography and they have a multi-scale spatial distribution \cite{dv03}. As such, they are a convenient and well-defined concept, unlike ``eddies''. They have also proved useful for understanding turbulent pair diffusion \cite{salazarcollins09}.

In this paper, we present a phenomenology based on the underlying topography of the fluctuating velocity field which relates the mean flow profile to the multiscale structure of stagnation points of the velocity fluctuations. We use Direct Numerical Simulations (DNS) of various fully developed incompressible turbulent channel flows to validate our new approach and propose a resulting new starting point for a new intermediate asymptotic analysis of the mean flow profile of turbulent channel/pipe flows. 

With the exception of the highest Reynolds number DNS channel flow data \cite{jimenezdnsdata} which we use towards the end of this paper, the Reynolds numbers considered here range between low to moderate (though, of course, always large enough for the flow to be turbulent). In terms of $\RE_\tau \equiv \tfrac{u_\tau \delta}{\nu} \equiv \delta / \delta_\nu$ where $u_\tau$ is the skin friction velocity, $\delta$ the channel half-width, $\nu$ the fluid's kinematic viscosity and the wall unit $\delta_\nu \equiv \nu / u_\tau$, the highest Reynolds number DNS data \cite{jimenezdnsdata} correspond to $\RE_\tau \simeq 950$ and $2000$. The Reynolds numbers of our own DNS data range between $\RE_\tau \simeq 110$ and $400$. This is too low for a direct assessment of the log-law but appears to be sufficient for the new approach to turbulent mean flow profiles which we propose here and which is based on stagnation points of the fluctuating velocity field.

The paper is organised as follows. In section II we describe our DNS of turbulent channel flows and in section III we present some of the conventional statistics which are obtained from our simulations. In section IV we introduce the stagnation point approach and its application to turbulent channel flows. The phenomenology and the mean flow properties implied by the results obtained from the application of this approach to our DNS are expounded in sections V and VI. Finally, some analysis of the highest Reynolds number DNS channel flow data \cite{jimenezdnsdata} is presented in section VII before summarising our conclusions in section VIII.
\section{DNS of turbulent channel flow}
We solve the non-dimensionalised incompressible Navier-Stokes equations in Cartesian coordinates
\begin{equation}
  \begin{gathered}
    \label{eq:NS}
    \grad \sdot \bm u = 0 \\
    \pd_t \bm u 
    + \frac{1}{2} \lt[ \grad(\bm u \ox \bm u) + (\bm u \sdot \grad)\bm u \rt]
    = - \grad p
    + \frac{1}{\RE_c} \bm{\Delta u}
  \end{gathered}
\end{equation}	
where $\RE_c \equiv U_c\delta / \nu$ is the Reynolds number based on $U_c \equiv \tfrac{3}{2} U_b$ and $U_b$ is the bulk velocity of the flow kept constant in time.

We use the code of Laizet \& Lamballais \cite{laizetlamballais09} where spatial derivatives are estimated using a sixth-order compact finite-difference scheme and equations (\ref{eq:NS}) are numerically integrated with a frational step method using a three-stage third-order Runge-Kutta scheme. The fractional step method projects the velocity field to a divergence free velocity field and the Poisson pressure equation is solved in Fourier space with a staggered grid for the pressure field.  The staggered grid for the pressure was used for numerical stability purposes as was the skew-symmetric implementation of the non-linear term in the Navier-Stokes equation (see eqn. (\ref{eq:NS})). The grid stretching technique maps an equally spaced co-ordinate in the computational space to a non-equally spaced co-ordinate in the physical space, in order to be able to use Fourier transforms in the inhomogeneous wall-normal direction \cite{laizetlamballais09, cainetal84}.

To simulate incompressible channel flow turbulence we adopted periodic boundary conditions for $\bm u \equiv (u,v,w)$ in the $x$ and $z$ directions except at the walls at $y = 0$ and $y = 2 \delta$ where the boundary conditions are either $\bm u = \bm 0$ or borrowed from studies of flow control schemes aimed at drag reduction \cite{xuetal07, minetal06}. The mean flow is in the $x$ direction (i.e. $\avg v = \avg w= 0$ but $\avg u \not = 0$, where the angle brackets $\avg.$, in this study, denote averages in the $x$ and $z$ homogeneous directions and time expect when, in section \ref{sec:stagnpoint} stagnation points of $\bm u - \avg{\bm u}$ are sought, in which case the average $\avg{\bm u}$ is only over $x$ and $z$) and the bulk velocity $U_b$ in that direction was kept at the same constant value at all times by a control procedure which adjusts the mean pressure gradient $- \dd\avg p / \dd x$ at each time step. The choice of $U_b$ is made in accordance with Dean's formula $\RE_\tau \simeq 0.119 \RE_{c}^{7/8}$ \cite{dean78} for a given choice of $\RE_{\tau}$.  The number of grid points $N_x$, $N_y$ and $N_z$ in the $x$, $y$ and $z$ directions as well as the domain sizes $L_x$ and $L_z$ in the $x$ and $z$ directions are given in Table 1. The domain size in the $y$ direction is of course $L_{y} = 2\delta$.

We use different near-wall forcings and boundary conditions at the walls  so as to demonstrate how our stagnation point approach accounts for the way that different wall actuations modify the mean flow profile. Specifically, we considered the following three control schemes:

(\emph i) $\bm u = \bm 0$ at the walls with forcing $\bm f(y) = (-A\sin(2\pi y/\Lambda)H(\Lambda - y),0,0)$ near the $y=0$ wall and similar forcing near the $y = 2\delta$ wall \cite{xuetal07} where $H$ is the Heaviside function, $A = 0.16U_c^2 / \delta \simeq u_\tau^2 / \delta_\nu$ and $\Lambda = 11\delta_\nu$ (case A1). The forcings are applied to the Navier-Stokes momentum equations (\ref{eq:NS}). This scheme corresponds to a steady wall-parallel forcing localised within eleven wall units from the walls and uniform in the direction parallel to them. This force field averages to zero if integrated across the channel; it decelerates the flow closest to the wall but accelerates it in the immediately adjacent thin region.

(\emph{ii}) $\bm u = (0,a\cos(\alpha(x - ct)),0)$ at the wall \cite{minetal06} with $a / U_c = 0.05$, $\alpha / \delta = 0.5$ and $c = - 2U_c$ (case A2). This boundary condition corresponds to a blowing-suction travelling wave on the wall.
 
(\emph{iii}) $\bm u = \bm 0$ at the walls and $v(x,y_d,z,t)$ replaced by $-v(x,y_d,z,t)$ at all $(x,z)$ points on the planes $y_d = 10\delta_\nu$ and $y_d = 2\delta - 10\delta_\nu$ (case A3). This corresponds to a computational control scheme whereby the normal velocity at a distance $y_d$ from the walls is made to change sign at every time step.

The numerical parameters of our computations are given in Table I.
\begin{table}[!h]
  \caption{(Color online) \label{dnsparameters}Parameters for the DNS of turbulent channel flow. The term ``Forcing" refers to wall or near-wall actuations.}
  \begin{ruledtabular}
    \begin{tabular}{ccccccc}
      \textbf{Case} & \textbf{Forcing} & $\bm{\RE_c}$ & $\bm{\RE_\tau}$ & $\bm{L_x}$ & $\bm{L_z}$ & $\bm{N_x \times N_y \times N_z}$ \\
      \hline
      A	& No 	& 4250 & 179 & $4\pi\delta$ & $4\pi\delta/3$ & $200 \times 129 \times 200$ \\
      A1& Yes \footnote{Xu \etal\ \cite{xuetal07}}
      	& 4250 & 114.4 & $4\pi\delta$ & $4\pi\delta/3$ & $200 \times 129 \times 200$ \\
      A2& Yes \footnote{Min \etal\ \cite{minetal06}}
       & 4250 & 222.3 & $4\pi\delta$ & $4\pi\delta/3$ & $200 \times 129 \times 200$ \\
      A3& Yes
       & 4250 & 141.6 & $4\pi\delta$ & $4\pi\delta/3$ & $200 \times 129 \times 200$ \\
      B	& No 	& 2400 & 109.5 & $4\pi\delta$ & $2\pi\delta$ & $100 \times 65 \times 100$ \\
      C	& No 	& 10400 & 392.6 & $2\pi\delta$ & $\pi\delta$ & $256 \times 257 \times 256$ \\
    \end{tabular}
  \end{ruledtabular}
\end{table}
Note that the net mass flux through the wall is zero in all the cases considered here and that the $y$-integrated momentum balance
\begin{equation}
  \label{eq:momentum}
  \nu\frac{\dd \avg u}{\dd y} - \avg{uv} = u_\tau^2(1 - \frac{y}{\delta}) 
\end{equation}
holds for all $y$ in all cases except with A1 forcing where it holds for $\Lambda < y < 2\delta - \Lambda$. All our non-forced computations have been validated against previously published databases \cite{mkmdnsdata, kasagidnsdata}.
\section{Conventional DNS results}
\begin{figure}[!h]
  \centering
  \includegraphics[width=8.5cm]{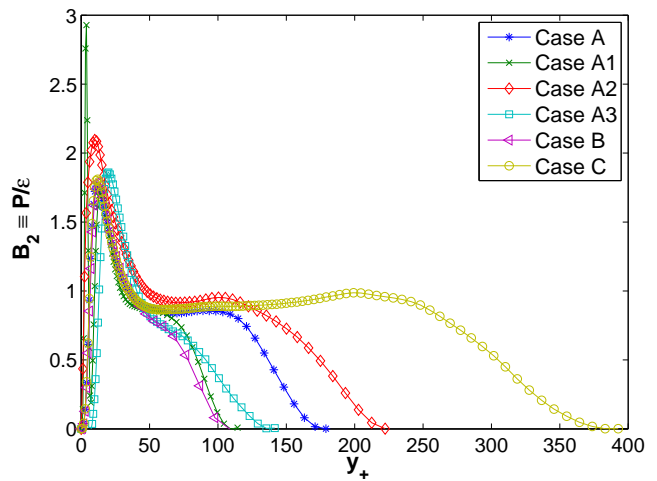}
  \caption{(Color online) Profile of the production to dissipation ratio. Note the
  existence of an approximate equilibrium layer which grows with $\RE_\tau$ and
  where production approximately balances dissipation.}
  \label{fig:b2}
\end{figure}

When $\RE_\tau \gg 1$ one might expect an intermediate region $\delta_\nu \ll y \ll \delta$ where production balances dissipation locally \cite{townsend61}, i.e. $- \avg{uv}\tfrac{\dd}{\dd y}\avg u \simeq \epsilon$. The idea of such an intermediate region is supported by our DNS results (see Fig. \ref{fig:b2}) which suggest that
\begin{equation}
  B_2 \equiv \mathcal P / \epsilon \equiv - \avg{uv}\tfrac{\dd}{\dd y}\avg u / \epsilon
\end{equation}
tends to 1 as $\RE_\tau \to \infty$ in this intermediate region (a recent paper \cite{brouwers07} proves this asymptotic result by assuming, however, that the mean flow has a logarithmic shape in the intermediate region) and that this region where this approximate balance holds also increases as $\RE_\tau$ increases. The slight discrepancy away from $B_2 \simeq 1$ at these moderate Reynolds numbers is well known and agrees with other previously published DNS results \cite{pope00}.

In this intermediate region, equation (\ref{eq:momentum}) implies $-\avg{uv} \simeq u_\tau^2$ as $y/\delta \rightarrow 0$ and $y_+ \equiv y/\delta_\nu \rightarrow \infty $, assuming that $\tfrac{\dd}{\dd \ln y_+} U_{+}$ (where $U_{+} \equiv \avg u / u_{\tau}$) does not increase faster than $y_+^p$ with $p \ge 1$ in this limit. It then follows that in this intermediate equilibrium region,
\begin{equation}
  \label{eq:loglaw}
  \epsilon \simeq \frac{u_\tau^3}{\kappa y} \; \text{implies}
  \; \frac{\dd \avg u}{\dd y} \simeq \frac{u_\tau}{\kappa y}
\end{equation}	
as $\RE_\tau \gg 1$. At finite Reynolds numbers the equation for the mean shear in (\ref{eq:loglaw}) should be replaced by $\frac{\dd \avg u}{\dd y} \simeq \frac{B_2}{B_3} \frac{u_\tau}{\kappa y}$ where 
\begin{equation}
  B_3 \equiv \frac{- \avg{uv}}{u_\tau^2}.
\end{equation}
Note that even though $B_2$ and $B_3$ may tend to 1 as $\RE_\tau \gg 1$, they are definitely different from 1 and even functions of $y_+$ and $y/\delta$ at finite values of $\RE_\tau$.

\begin{figure}[!h]
  \centering
  \includegraphics[width=8.5cm]{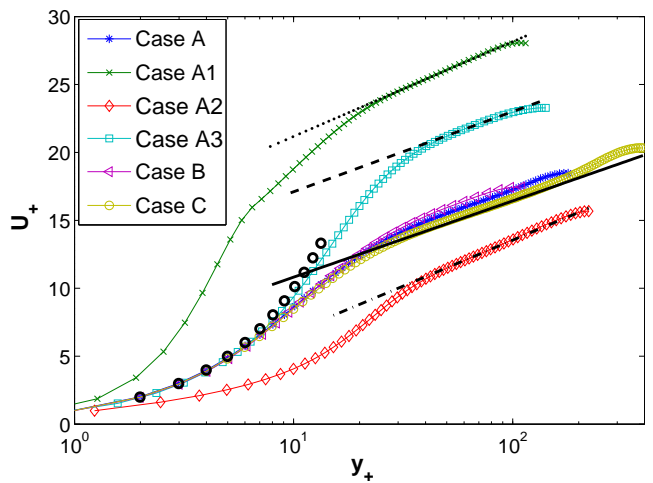}
  \caption{(Color online) Mean velocity profiles. For comparison we also plot best
  log-law fits. $\circ$: $U_+ = y_+$, $\dotsi$: $U_+ = \frac{1}{0.33}y_+ + 14.2$, -- $\sdot$ -- : $U_+ = \frac{1}{0.34}y_+ + 0.0$, - - -: $U_+ = \frac{1}{0.39}y_+ + 11.2$, -----: $U_+ = \frac{1}{0.41}y_+ + 5.2$}
  \label{fig:meanflow}
\end{figure}

The mean flow profiles show clear impacts of the control schemes on the mean flow (see Fig. \ref{fig:meanflow}). For our various control schemes at the same $\RE_c = 4250$, the skin friction decreases as a result of both case A1 and A3 but increases when the control scheme A2 is applied (see Table I). This observation agrees with Fig. \ref{fig:meanflow} where mean flow values for cases A1 and A3 are higher than for case A (no control scheme), and mean flow values are lower for case A2 than for case A. 

\begin{figure}[!h]
  \centering
  \includegraphics[width=8.5cm]{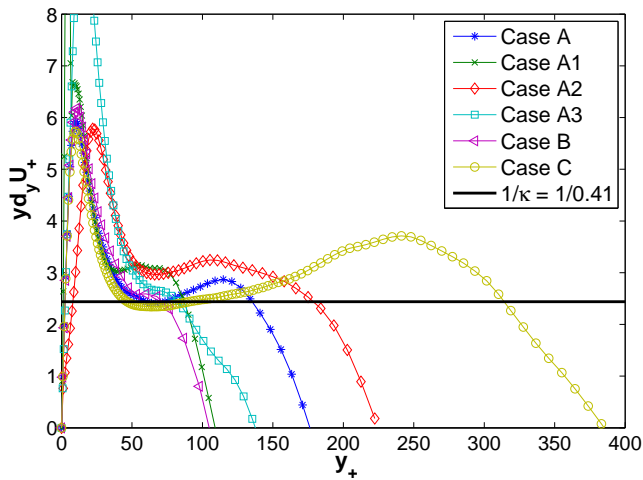}
  \caption{(Color online) The inverse von K\'arm\'an coefficient $\equiv y \tfrac{\dd}{\dd y}U_+$ versus $y_+$. Taking the definition of $\kappa$ to be given by the left-hand equation in (\ref{eq:loglaw}) it is really $B_2 / (B_3 \kappa)$ which is plotted against $y_+$. The effects of the various near-wall actuations are significant.}
  \label{fig:invkappa}
\end{figure}

\begin{figure}[!h]
  \centering
  \includegraphics[width=8.5cm]{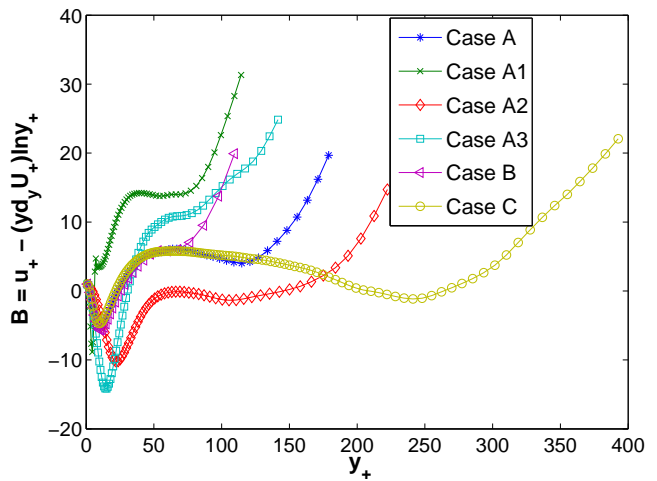}
  \caption{(Color online) $B \equiv U_+ - \lt( y \tfrac{\dd}{\dd y}U_+ \rt) \log y_+$ versus $y_+$ for the six different DNS cases in Table I.}
  \label{fig:beta}
\end{figure}

With reference to the log-law scaling $U_+ = \frac{1}{\kappa} \log y_+ + B$ which results from integration of (\ref{eq:loglaw}) if $1/\kappa$ is independent of $y$, we plot the coefficient $y \tfrac{\dd}{\dd y}U_+$ versus $y_+$ in Fig. \ref{fig:invkappa} and the coefficient $B \equiv U_+ - \lt( y \frac{\dd}{\dd y}U_+ \rt) \log y_+$ versus $y_+$ in Fig. \ref{fig:beta} for all the six different DNS cases of Table I. Note that $y \tfrac{\dd}{\dd y}U_+$ is usually refered to as $1/\kappa$ but is in fact $B_2 / (B_3 \kappa)$ in the present context where $\kappa$ is defined by the left-hand side equation in (\ref{eq:loglaw}). It is only if $B_2$ and $B_3$ both equal 1 in the equilibrium layer, as may be the case when $\RE_\tau \gg 1$, that $\frac{\dd}{\dd y}\avg u \simeq \frac{B_2}{B_3} \frac{u_\tau}{\kappa y}$ yields $\frac{\dd}{\dd y}\avg u \simeq \frac{u_\tau}{\kappa y}$ and that $y \tfrac{\dd}{\dd y} U_+$ becomes $1/\kappa$ in the equilibrium layer.

The values of $B$ are affected by the various control schemes (see Fig. \ref{fig:beta}) in a way consistent with the observations made two paragraphs earlier (higher values of $B$ for cases A1 and A3 than for A and lower for case A2). However, it is hard to conclude on the validity of the log-law from these results and in particular from the plot in Fig. \ref{fig:invkappa} which clearly shows a significant dependence on near-wall conditions, $\RE_\tau$ and $y_+$. It may be that the log-law is not valid at all or it may be that the log-law is not valid unless the Reynolds number is sufficiently high, definitely higher than the Reynolds numbers of our simulations.
\section{The stagnation point approach}
\label{sec:stagnpoint}
As our direct DNS study of the mean flow equation in (\ref{eq:loglaw}) does not yield clear results, we chose instead to investigate the validity of the dissipation equation in (\ref{eq:loglaw}). For this we make use of the stagnation point approach which has been used recently to show how the number density of stagnation points in high Reynolds number Homogeneous Isotropic Turbulence (HIT) determines salient properties of turbulent pair diffusion \cite{salazarcollins09} and kinetic energy dissipation rate per unit mass \cite{mv08, gv09}. In particular, a generalised Rice theorem was recently proved \cite{gv09} for high Reynolds number HIT which states that \textit{the Taylor microscale is proportional to the average distance between neighboring stagnation points}. This average distance is defined as the $-1/d$ power of the number density of stagnation points which are points in the $d$-dimensional space of the flow where the turbulent fluctuation velocity is zero.

The generalised Rice theorem \cite{gv09} for high Reynolds number HIT holds under two main assumptions: (\emph i) statistical independence between large and small scales and (\emph{ii}) absence of small-scale intermittency effects. The question which arises in the context of the present work is whether it also holds in some region of turbulent channel flows. For other cases where measures and concepts from HIT impact wall-bounded turbulence please refer to \cite{lundgren07, gioiachakraborty06}.

\begin{figure}[!h]
  \centering
  \includegraphics[width=8.5cm]{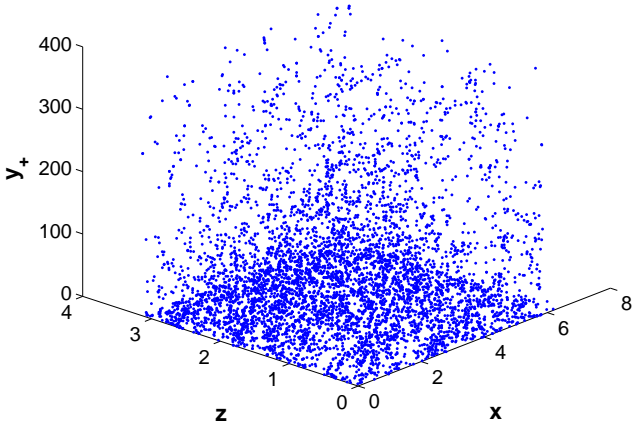}
  \caption{(Color online) Points where $\bm u' \equiv \bm u - \avg {\bm u}= 0$ for case C at a given instant in time}
  \label{fig:stgnpts}
\end{figure}

To obtain some insight into this question by way of our DNS, we consider stagnation points of the turbulent fluctuation velocity field ${\bm u'} \equiv {\bm u} - \avg {\bm u}$, i.e. points where all components of the velocity fluctuations around the local mean flow are zero. A 3D plot of these points for an instant in time in our DNS channel is presented in Fig. \ref{fig:stgnpts}. We use fourth-order Lagrangian interpolation and the Newton-Raphson method to locate these points. Details on how these points are found are given in the Appendix \ref{app:stgnptsmethod}.

\begin{figure}[!h]
  \centering
  \includegraphics[width=8.5cm]{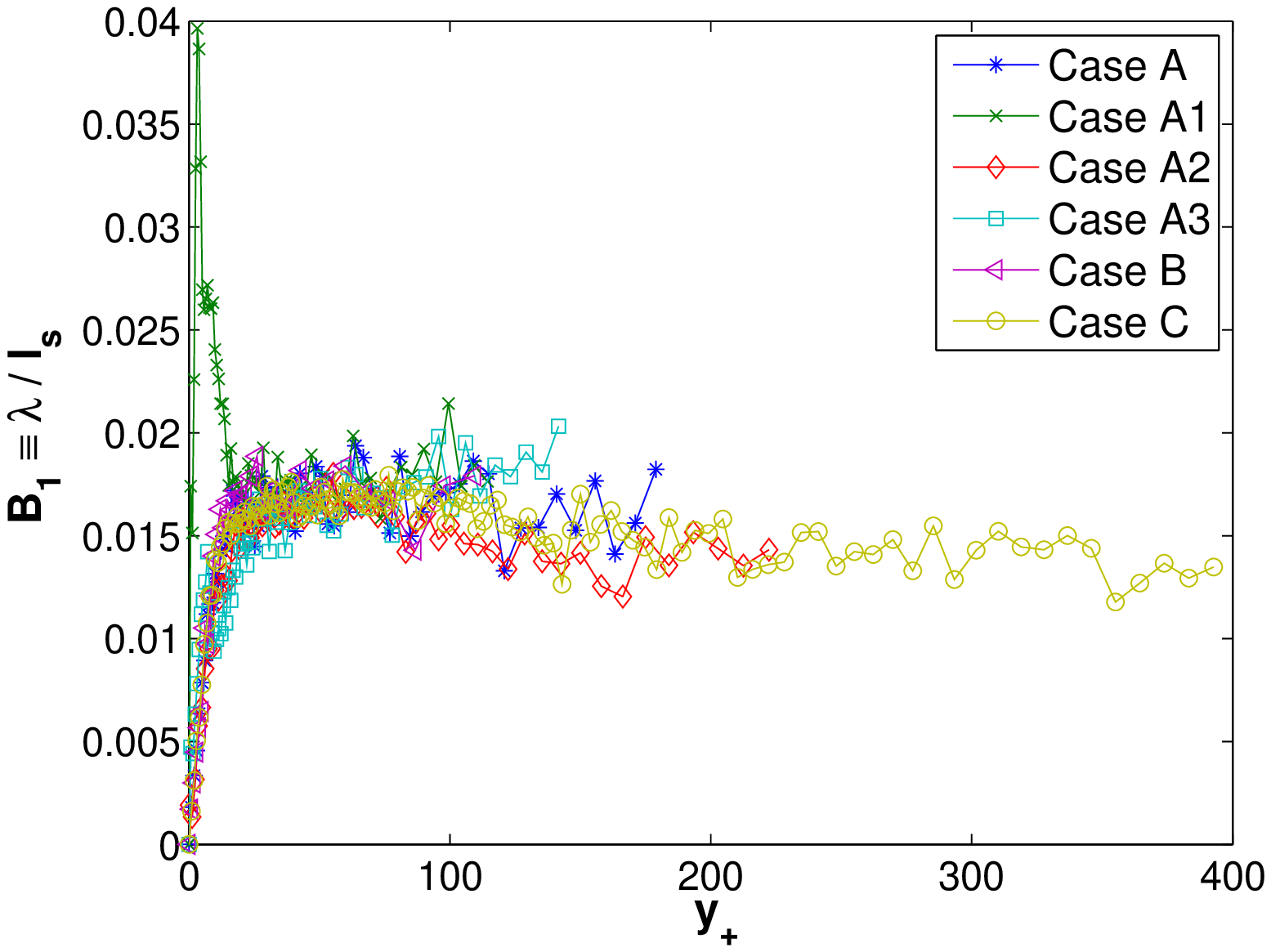}
  \includegraphics[width=8.5cm]{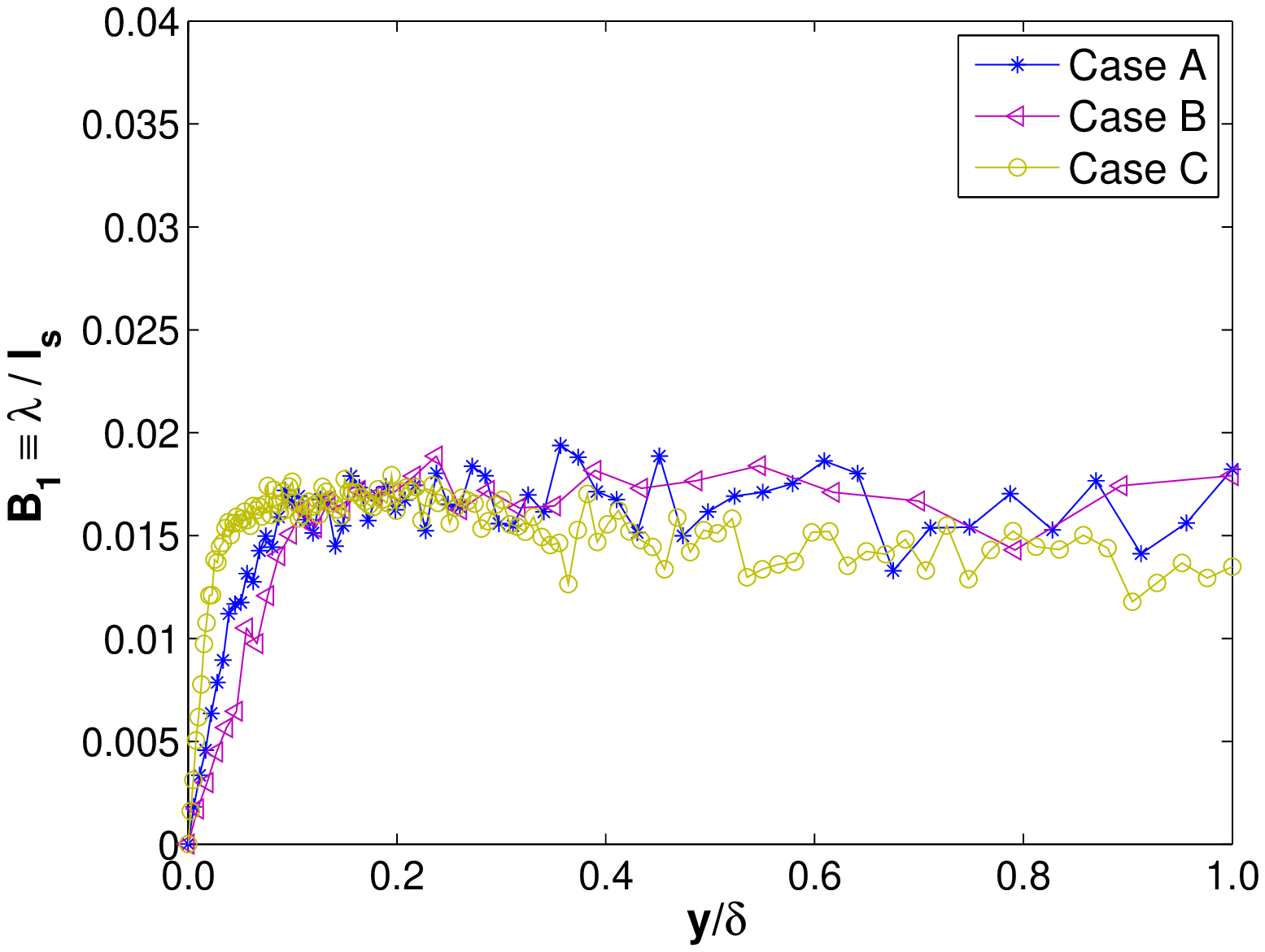}
  \includegraphics[width=8.3cm]{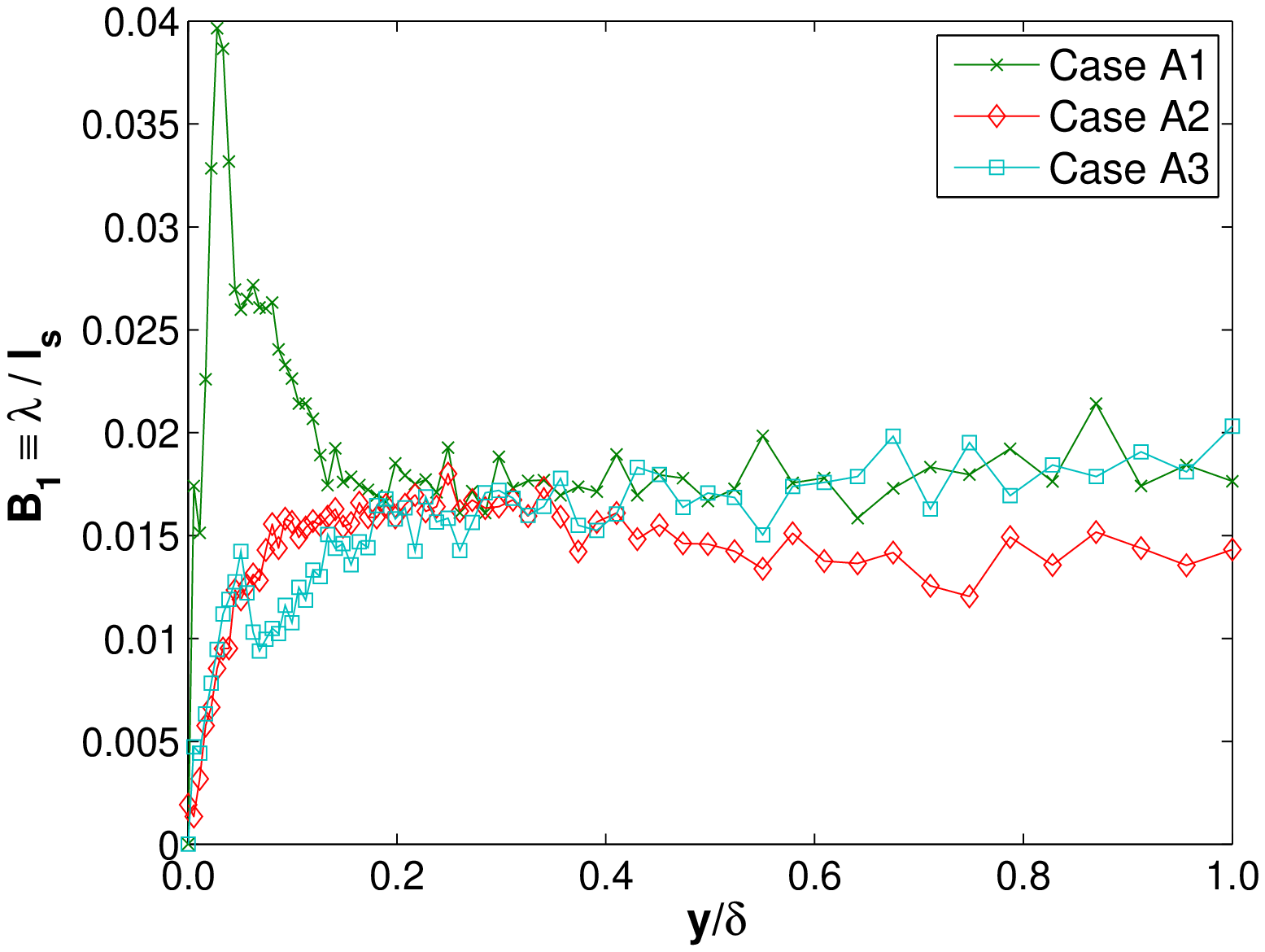}
  \caption{(Color online) Support for the generalised Rice theorem as a meaningful
  approximation in turbulent channel flows with various Reynolds
  numbers and wall actuations (see Table I). (a) $B_1$ as a function of $y_+$, (b) $B_1$ as a function of $y/\delta$ with no wall-forcings and (c) $B_1$ as a function of $y/\delta$ with wall-forcings.}
  \label{fig:b1}
\end{figure}

We define $N_s(y_+)$ to be the total number of these stagnation points in a thin slab parallel to and at a distance $y$ from the channel's $y = 0$ wall (this number can also be defined with respect to the $y = 2\delta$ wall, but here we only consider $0 \le y \le \delta$ without loss of generality). The dimensions of this slab are $L_x \times \delta_y \times L_z$ with slab thickness $\delta_y \sim \delta_\nu$. The average distance between stagnation points at a height $y$ from the wall is $\ell_s \equiv \sqrt{\tfrac{L_x L_z}{N_s}}$. A Taylor microscale $\lambda(y)$ can be defined from $\epsilon(y) = 2\nu \avg{s_{ij}s_{ij}} \equiv \tfrac{\nu}{3}\tfrac{2E}{\lambda^2}$ where $s_{ij}$ is the fluctuating velocity's strain rate tensor and $E(y) \equiv \tfrac{1}{2}\avg{|\bm u|^2}$. The question raised is whether a region of turbulent channel flow exists for $\RE_\tau \gg 1$ where
\begin{equation}
  \label{eq:ricethm}
  \lambda(y) = B_1 \ell_s(y)
\end{equation}
with $B_1$ independent of $y$ and Reynolds number. The answer provided by our DNS is that $B_1$ is indeed approximately constant over an intermediate range $\delta_\nu \ll y \lesssim \delta$, but not perfectly so as our plots in Fig. \ref{fig:b1} attest to. It is worth noting that this constancy of $B_1$ appears to be better defined for cases A, B and C where there is no wall or near-wall actuation. Hence, a small discrepancy away from $B_1 = Const$ may be achieved as a result of those different wall-forcings. However, part of the even smaller discrepancy in cases A, B and C might be accountable to neglected small-scale intermittency effects which, in the case of high Reynolds number HIT, are known to manifest themselves as a weak Reynolds number dependence on $B_1$ \cite{mv08}. In the case of wall-bounded turbulence, small-scale intermittency effects could therefore manifest themselves as a weak dependence of $B_1$ on local Reynolds number $y_+ = y / \delta_\nu$ (see also \cite{mehrafarinpourtolami08}). However, we leave this refinement for future studies.

Using (\ref{eq:ricethm}) and $\ell_s \equiv \sqrt{\tfrac{L_x L_z}{N_s}}$, we write 
\begin{equation}
  \label{eq:dissipation}
  \epsilon = \frac{\nu}{3}\frac{2E}{\lambda^2} =
  \frac{\nu}{3}\frac{2E}{B_1^2 \ell_s^2} =
  \frac{\nu}{3}\frac{2E}{B_1^2 L_x L_z}N_s =
  \frac{\nu}{3}\frac{2E}{B_1^2}\delta_\nu n_s
\end{equation}
where we have introduced the number density of stagnation points $n_s \equiv N_s / (L_x L_z \delta_\nu)$. Combining this last equation with $-\avg{uv}\tfrac{\dd}{\dd y}\avg u = B_2 \epsilon$ and using $\tfrac{\dd}{\dd y}\avg u = \tfrac{u_\tau}{\kappa y}$ as well as $C \equiv - \frac{2E}{3\avg{uv}}$, we obtain 
\begin{equation}
  \label{eq:-1law}
  n_s = \frac{C_s}{\delta_\nu^3}y_+^{-1}
\end{equation} 
where $C_s$ is given by
\begin{equation}
  \label{eq:invkappacs}
  C_s = \frac{B_1^2}{\kappa B_2 C}. 
\end{equation}

\begin{figure}[!h]
  \centering
  \includegraphics[width=8.5cm]{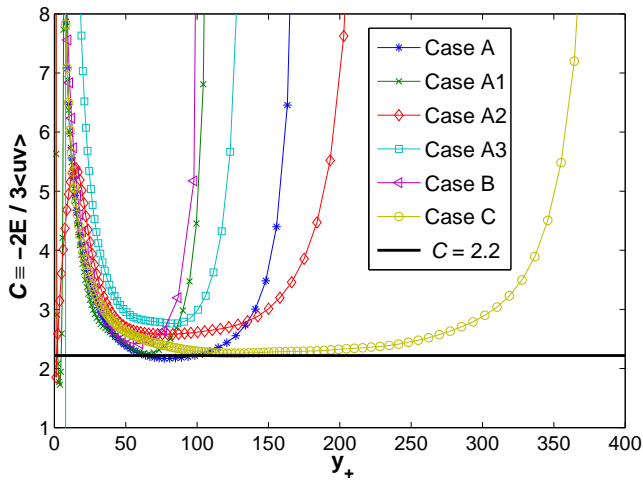}
  \caption{(Color online) $C$ as a function of $y_+$ for different Reynolds numbers
  and different wall-actuation cases (see Table I).}
  \label{fig:c}
\end{figure}

\begin{figure}[!h]
  \centering
  \includegraphics[width=8.5cm]{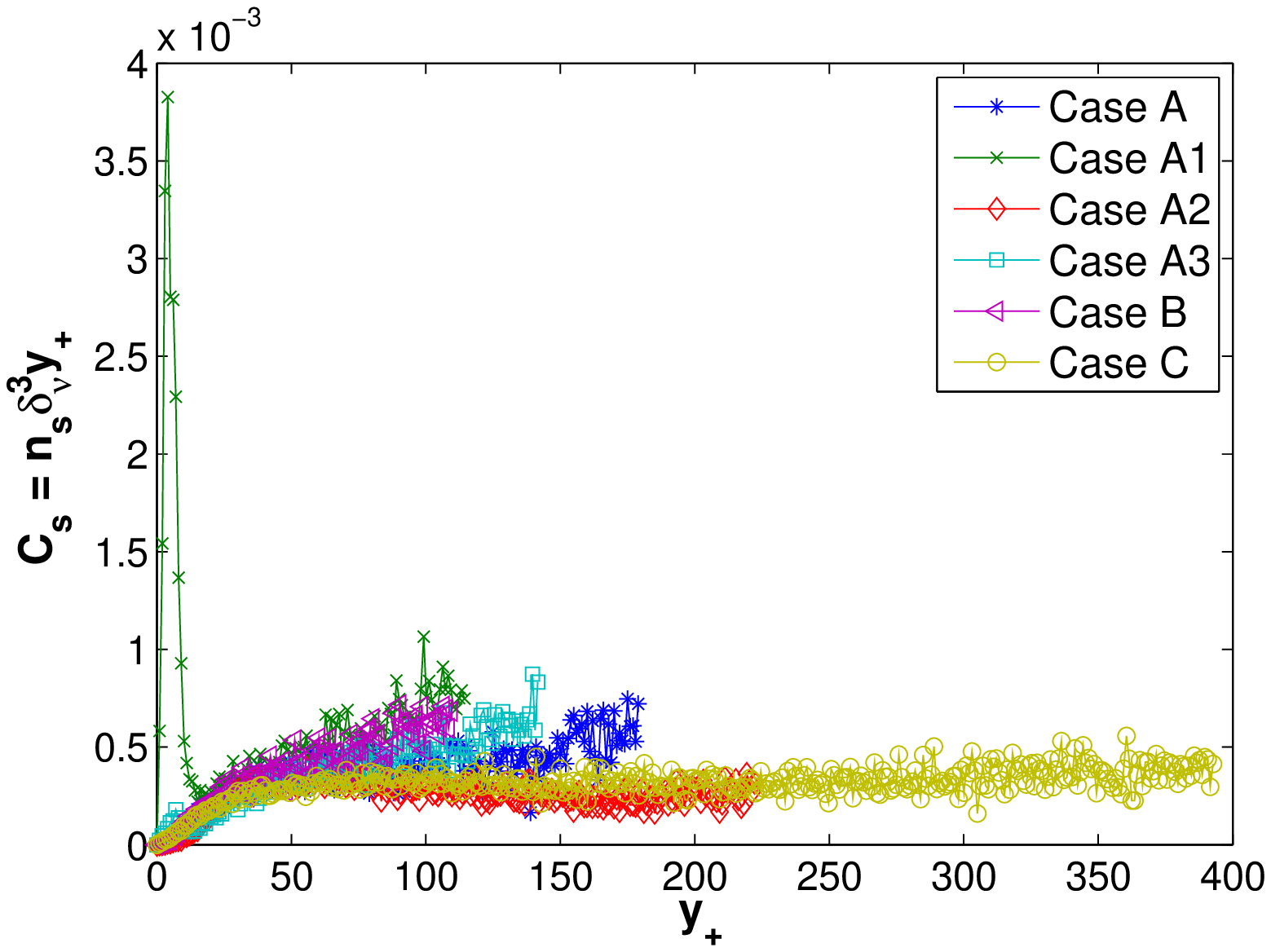}
  \includegraphics[width=8.3cm]{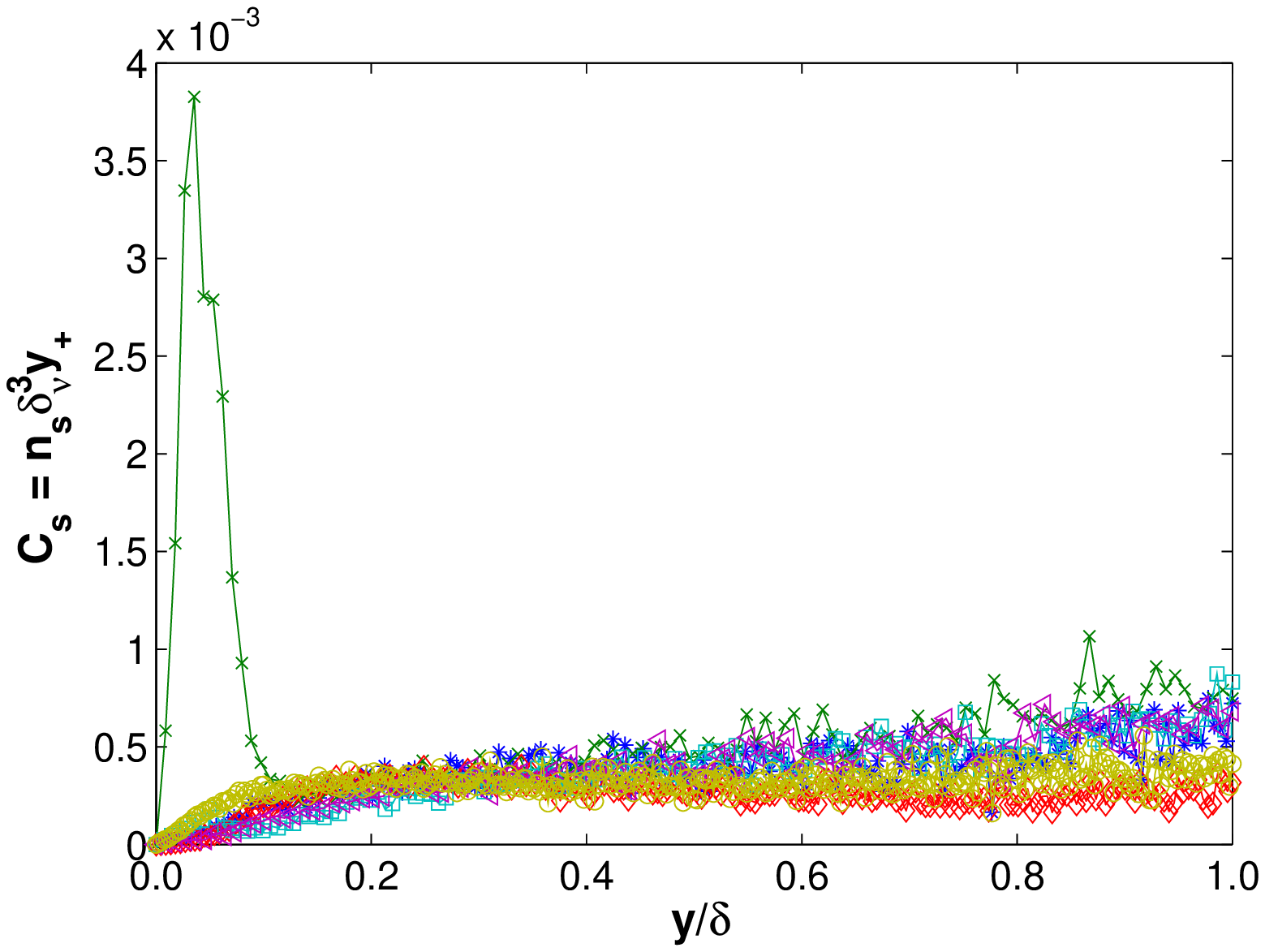}
  \caption{(Color online) Normalised number of turbulent velocity stagnation points
  for different Reynolds numbers and different wall-actuation cases
  (see Table I). (a) $C_{s}$ versus $y_+$; (b) $C_{s}$ versus
  $y/\delta$.}
  \label{fig:cs}
\end{figure}

The classical claims \cite{pope00} are that $\kappa \simeq 0.4$, $C \simeq 2$ and $B_{2} \simeq 1$ in the intermediate range $1 \ll y_+ \ll \RE_\tau$ as $\RE_\tau \to \infty$. These claims therefore imply that $C_s$ should also be a constant in that same range and limit provided $B_1$ is. Whilst, as we have seen, $B_1$ is not too far from being constant in the range $\delta_\nu \ll y \lesssim \delta$, $\kappa$ and $C$ are significantly far from constant in this range (see Fig. \ref{fig:invkappa} and Fig. \ref{fig:c}). Even so, our DNS evidence (see Fig. \ref{fig:cs}) suggests that $C_s$ tends to a well-defined constant in the range $\delta_\nu \ll y \lesssim \delta$ as $\RE_\tau$ increases. Remarkably, this condition on $\RE_\tau$ for the constancies of $C_s$ and $B_1$ seems to require as little as $\RE_\tau$ exceeding a few hundred. It is equally remarkable that our calculation of $N_s$, which underpins $B_1$ and $C_s$, has involved an average over a number of time-samples that is two orders of magnitude smaller than for the time average required to statistically converge $\avg u$, $\avg{uv}$, $E$ and $\epsilon$.

The constancy of $C_s$ in the range $\delta_\nu \ll y \lesssim \delta$ implies that, in that range, the number density $n_s$ of stagnation points decreases with distance from the wall as $y_+^{-1}$. This is in qualitative agreement with Fig. \ref{fig:stgnpts} which shows the stagnation points to be increasingly denser as the wall is aproached.

The constant $C_s$ can be interpreted as representing the number of turbulent velocity stagnation points within a cube of side-length equal to a few multiples of $\delta_{\nu}$ (see (\ref{eq:-1law})) placed where $y$ equals a few multiples of $10\delta_\nu$ as seen in Fig. (\ref{fig:cs}a). This is the lower end of the range where $n_s \sim y_+^{-1}$ and seems to be where the upper edge of the buffer layer is usually claimed to lie \cite{pope00}.

Equations (\ref{eq:-1law}) and (\ref{eq:invkappacs}) have been derived by assuming well-defined constant values of $\kappa$, $B_2$, $C$ and $B_1$. However, our DNS results show that, at the Reynolds numbers considered, $B_1$ and $C_s$ are indeed constant but $\kappa$, $B_2$ and $C$ are clearly not. Equations (\ref{eq:ricethm}) and (\ref{eq:-1law}) with constant dimensionless values of $B_1$ and $C_s$ seem to be more broadly valid than the assumptions under which we derived equation (\ref{eq:-1law}). Therefore, in the next section we explore the phenomenology behind the new equations (\ref{eq:ricethm}) and (\ref{eq:-1law}) and the constant values of $B_1$ and $C_s$ in the range $\delta_\nu \ll y \lesssim \delta$ and in section \ref{sec:vi}, we go one step further and derive the consequences of the constancies of $B_1$ and $C_s$ on the mean flow profile without assuming well-defined constant values of $\kappa$, $B_2$ and $C$.
\section{Phenomenology}
One interpretation of the constancy of $B_1$ can be obtained by considering the eddy turnover time $\tau$ which is defined by $\epsilon = E / \tau$. Combined with the equation $\epsilon = \frac{\nu}{3} \frac{2E}{\lambda^2}$ which defines $\lambda$, one obtains $3 \lambda^2 = 2 \nu \tau$. Using (\ref{eq:ricethm}), $B_1 = Const$ is then equivalent to
\begin{equation}
  \label{eq:turnovertime}
  \frac{\ell_s^2}{\nu} \sim \tau 
\end{equation}
which indicates that in the equilibrium layer, the time it takes for viscous diffusion to spread over neighbouring stagnation points is the same proportion of the eddy turnover time at all locations and all Reynolds numbers. In high Reynolds number turbulence, the turnover time is also the time it takes for the energy to cascade to the smallest scales.

For an interpretation of the constancy of $C_s$ note first that (\ref{eq:-1law}) and $\ell_s = \sqrt{\frac{L_xL_z}{N_s}} = (n_s \delta_\nu)^{-1/2}$ imply $\ell_s^2 = C_s^{-1} \delta_\nu y$. From $\epsilon = \frac{\nu}{3}\frac{2E}{\lambda^2}$ and $B_1 = \lambda / \ell_s$ it then follows that
\begin{equation}
  \label{eq:dissipationlaw}
  \epsilon = \frac{2}{3} \frac{E u_\tau}{\kappa_s y}
\end{equation}
with 
\begin{equation}
  \label{eq:kappastar}
  \kappa_s \equiv \frac{B_1^2}{C_s}.
\end{equation}
The meaning of $C_s$ and $B_1$ constant is therefore, using (\ref{eq:dissipationlaw}), that the eddy turnover time $\tau \equiv E/\epsilon$ is proportional to $y/u_{\tau}$ throughout the range where they are constant. The constant of proportionality is $3\kappa_s /2$ where $\kappa_s$ is determined by the stagnation point coefficients $B_1$ and $C_s$ and is constant if they are constant. We refer to $\kappa_s$ as the stagnation point von K\'arm\'an coefficient.

Note that, in the present context, equation (\ref{eq:dissipationlaw}) replaces the usual $\epsilon = u_\tau^3 / \kappa y$ \cite{pope00}, and that these two equations reduce to the same one only if and where $E \sim u_\tau^2$ independently of $y_+$ and $\RE_\tau$.
\section{The mean flow profile in the equilibrium layer}
\label{sec:vi}
In this section we spell out the consequences of the constancies of $B_1$ and $C_s$ on the mean flow profile.

In the equilibrium layer the expectation is that $B_2 \rightarrow 1$ in the limit $\RE_\tau \rightarrow \infty$. This means that $-\avg{uv}\tfrac{\dd}{\dd y}\avg u = B_2 \epsilon$ may be replaced by $- \avg{uv}\tfrac{\dd}{\dd y}\avg u = \epsilon$ in the equilibrium layer. The constancy of $B_1$ and $C_s$ in this same limit implies a constant $\kappa_s = B_1^2 / C_s$ in $\epsilon = \tfrac{2}{3}E_+ \tfrac{u_\tau^3}{\kappa_s y}$ where $E_+ \equiv E / u_\tau^2$. It then follows that $-\avg{uv}\tfrac{\dd}{\dd y}\avg u = \epsilon = \frac{2}{3} E_+ \tfrac{u_\tau^3}{\kappa_s y}$. In turbulent channel/pipe flows where we can have some mathematical confidence that, as $\RE_\tau \rightarrow \infty$, $-\avg{uv} \rightarrow u_\tau^2$ in an intermediate layer $\delta_\nu \ll y \ll \delta$, it finally follows that
\begin{equation}
  \label{eq:newmeanflow}
  \frac{\dd \avg u}{\dd y} \simeq \frac{2}{3} E_+ \frac{u_\tau}{\kappa_s y}
\end{equation}
in that same layer and limit. At finite Reynolds numbers this new equation (\ref{eq:newmeanflow}) should be replaced by $\frac{\dd}{\dd y}\avg u \simeq \frac{2}{3} \frac{B_2}{B_3} E_+ \frac{u_\tau}{\kappa_s y}$ and account should be taken of the fact that $B_2$, $B_3$ and $\kappa_s$ all have their own, potentially different, rates of convergence towards their high Reynolds number asymptotic constant values.

\begin{figure}[!h]
  \centering
  \includegraphics[width=8.5cm]{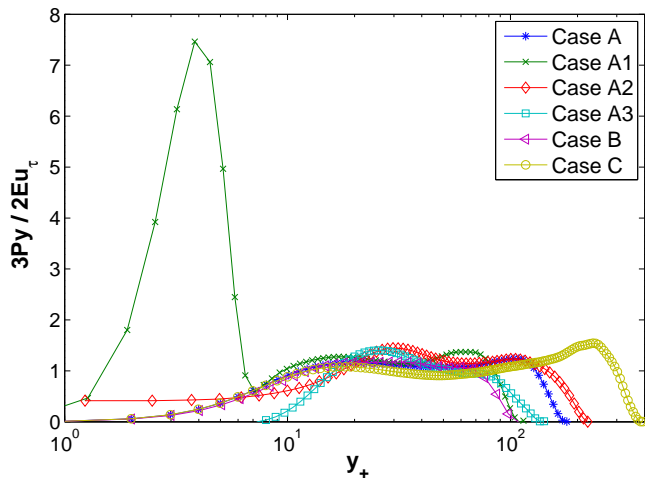}
  \caption{(Color online) Linear-log plots of $\frac{3}{2}\frac{\mathcal P}{E_+} \frac{y}{u_\tau^3}$ versus $y_+$ for different Reynolds numbers and different wall-actuation cases (see Table I). This is the same as $B_{2}/\kappa_s$ versus $y_+$ because of (\ref{eq:dissipationlaw}) and $\mathcal P = B_{2} \epsilon$.}
  \label{fig:invkappastar-eps}
\end{figure}

\begin{figure}[!h]
  \centering
  \includegraphics[width=8.5cm]{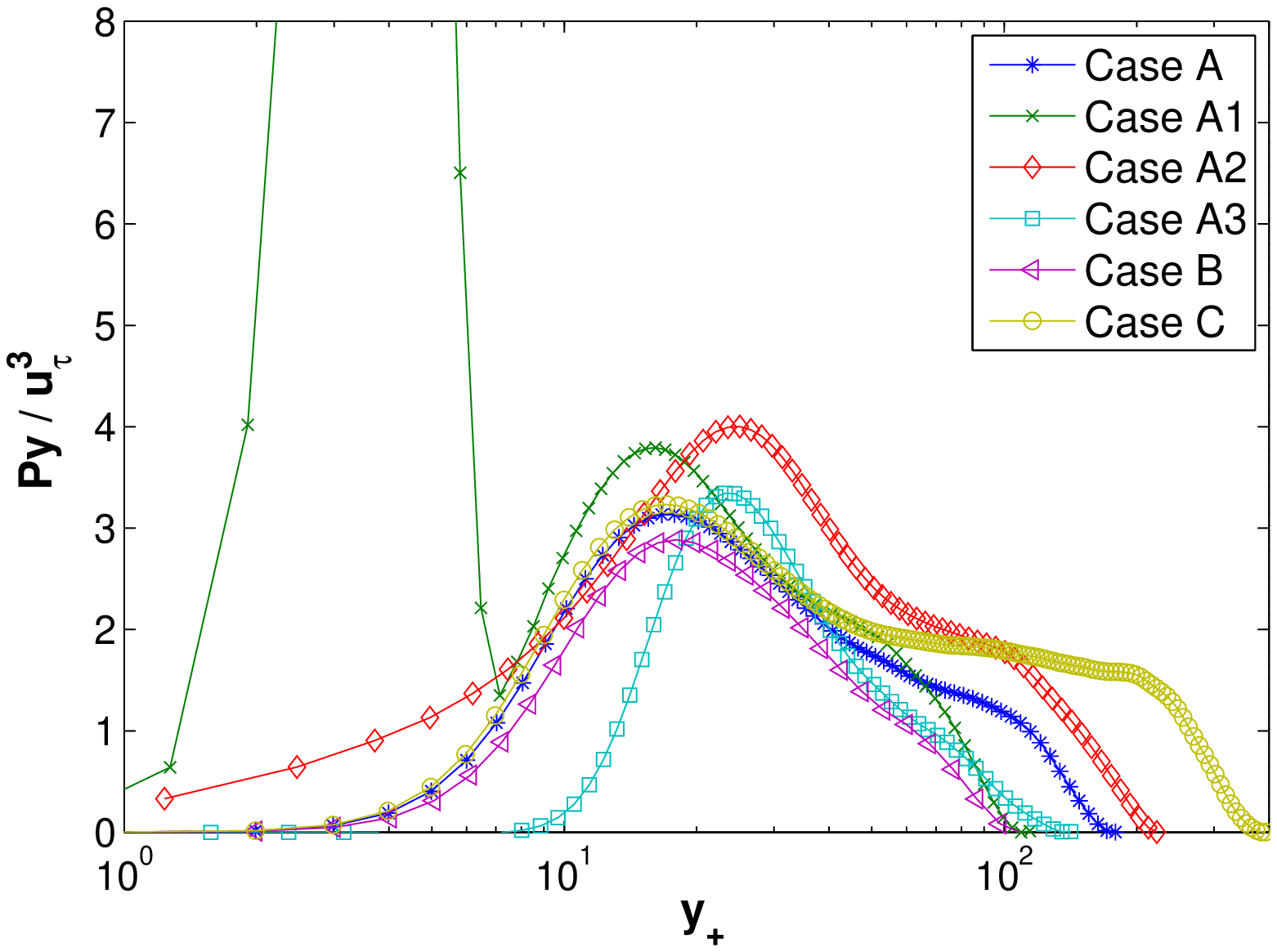}
  \caption{(Color online) Linear-log plots of $\mathcal P y / u_\tau^3$ versus $y_+$ for different Reynolds numbers and different wall-actuation cases (see Table I). This is the same as $B_2 / \kappa$ versus $y_+$ because of (\ref{eq:loglaw}) and $\mathcal P = B_2 \epsilon$.}
  \label{fig:epsclassic}
\end{figure}

An important step taken in deriving both (\ref{eq:loglaw}) and (\ref{eq:newmeanflow}) has been the local high Reynolds number balance $\mathcal P \simeq \epsilon$ in the equilibrium layer. In terms of the classical assumption $\epsilon \simeq u_\tau^3 / \kappa y$, $\mathcal P = B_2 \epsilon$ implies that $\mathcal P y/u_\tau^3$ should equal $B_2 / \kappa$ which should be constant in the equilibrium layer as a result of the balance between $\mathcal P$ and $\epsilon$. In terms of the new formula (\ref{eq:dissipationlaw}), $\mathcal P = B_2 \epsilon$ implies that $\frac{3}{2} \mathcal P y / (E_+ u_\tau^3)$ should equal $B_2 / \kappa_s$ and the balance between $\mathcal P$ and $\epsilon$ means that it should be $B_2 / \kappa_s$ rather than $B_2 / \kappa$ which is constant in the equilibrium layer. The main difference is the presence of $E$ in (\ref{eq:dissipationlaw}). In Fig. \ref{fig:invkappastar-eps} we plot our DNS results for $\frac{3}{2} \mathcal P y / (E_+ u_\tau^3)$ versus $y_+$ and in Fig. \ref{fig:epsclassic} our DNS results for $\mathcal P y / u_\tau^3$. It is clear that the collapse between the different Reynolds number and wall-actuation data is far worse and the $y$-dependence in the equilibrium layer far stronger for $B_2 / \kappa$ than for $B_2 / \kappa_s$. 

\begin{figure}[!h]
  \centering
  \includegraphics[width=8.5cm]{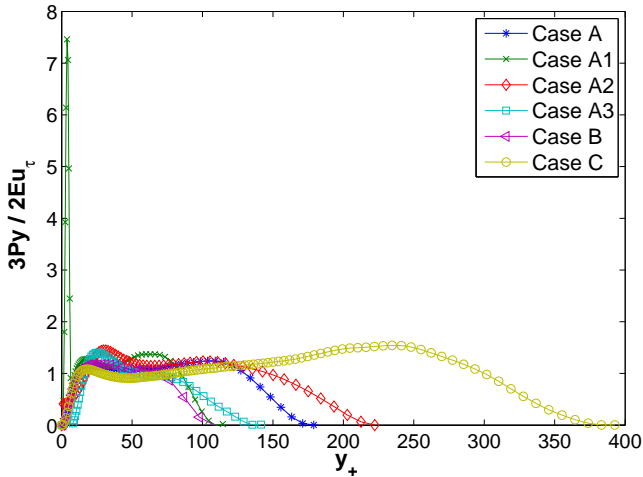}
  \caption{(Color online) Linear-linear plots of $B_2 / \kappa_s$ versus $y_+$ for different Reynolds numbers and different wall-actuation cases (see Table I).}
  \label{fig:invkappa-prod}
\end{figure}

These DNS results are for Reynolds numbers which are not very large; yet the high-Reynolds number constancy of $B_2 / \kappa_s$ in the equilibrium layer seems already not exceedingly far from being reached whereas no such indication is shown in the plot of $B_2 / \kappa$ versus $y_+$. Fig. \ref{fig:invkappa-prod} is a linear-linear replot of Fig. \ref{fig:invkappastar-eps} for easier comparison with Fig. \ref{fig:invkappa}.

\begin{figure}[!h]
   \centering
\includegraphics[width=8.5cm]{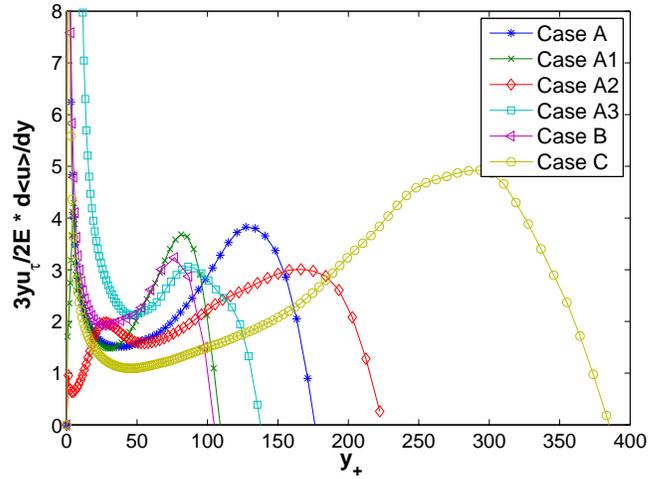}
   \caption{(Color online) Plots of $\frac{3}{2} \frac{y}{E_+u_\tau} \frac{\dd}{\dd y}\avg u$ versus $y_+$ for different Reynolds numbers and different wall actuations (see Table I).}
   \label{fig:eq11}
\end{figure}

\begin{figure}[!h]
 \centering
 \includegraphics[width=8.5cm]{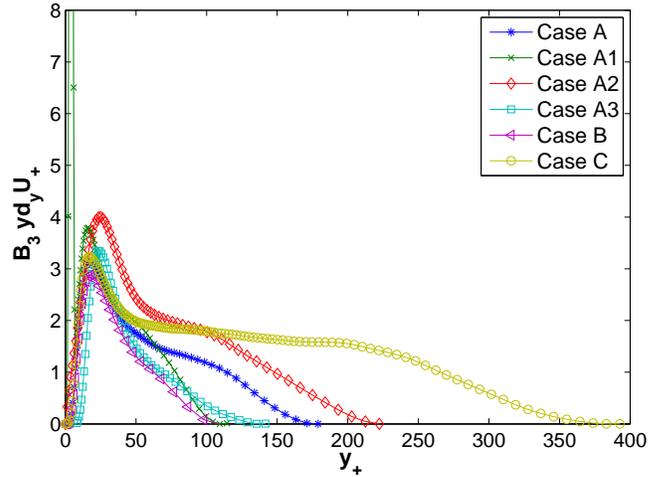}
 \caption{(Color online) Linear-linear plots of $B_3 \frac{y}{u_\tau} \frac{\dd}{\dd y}\avg u$ versus $y_+$ for different Reynolds numbers and different wall actuations (see Table I). These are effectively plots of $B_2 / \kappa$ to be compared with the similarly plotted $B_2 / \kappa_s$ in Fig. \ref{fig:invkappa-prod}.}
   \label{fig:comp}
\end{figure}

From (\ref{eq:newmeanflow}), direct plots of $\frac{3}{2} \frac{y}{E_+ u_\tau} \frac{\dd}{\dd y}\avg u$ should give $1/\kappa_s$ in the equilibrium layer when $\RE_\tau \to \infty$ and $B_2 / (B_3 \kappa_s)$ in that layer at finite Reynolds numbers. These plots for each of the Table I cases are given in Fig. \ref{fig:eq11}. They do not compare favourably with the plots of $\frac{y}{u_\tau} \frac{\dd}{\dd y}\avg u$, effectively plots of $B_2 / (B_3 \kappa)$, in Fig. \ref{fig:invkappa}. However, this does not mean that in the limit $\RE_\tau \to \infty$, (\ref{eq:loglaw}) is better than (\ref{eq:newmeanflow}) in the equilibrium layer. The facts that $C_s$ and $B_1$ are approximately constant in the range $\delta_\nu \ll y \lesssim \delta$ and that $B_2 / \kappa$ is much less collapsed and less constant along $y$ than $B_2 / \kappa_s$ at Table I's Reynolds numbers (see Fig. \ref{fig:comp} and compare it with Fig. \ref{fig:invkappa-prod}) suggest that the strong $y$ and $\RE_\tau$ dependencies of $B_3$ partly cancel those of $B_2 / \kappa$ at those Reynolds numbers. As the Reynolds number is increased to the point where $B_3$ reaches its asymptotic value 1 then this cancellation will either disappear if $B_2 / \kappa$ does not tend to a constant or will remain if it does. In the specific context of the present stagnation point approach, the choice between these two scenarios will depend on the high-Reynolds number scalings of the kinetic energy $E$.

According to classical similarity scalings, as $\RE_\tau \to \infty$, $E \sim u_\tau^2$ independently of $y$, $\delta$ and $\nu$ in the equilibrium range $\delta_\nu \ll y \ll \delta$. If this is true, then (\ref{eq:loglaw}) and the log-law are recovered from (\ref{eq:dissipationlaw}) and (\ref{eq:newmeanflow}) with a von K\'arm\'an coefficient $\kappa$ proportional to $\kappa_s$. However, Townsend's \cite{townsend76} idea of inactive motions would suggest that $E$ does not scale as $u_\tau^{2}$ in the equilibrium layer when $\RE_\tau \to \infty$. If this is the case, then (\ref{eq:newmeanflow}) does not yield (\ref{eq:loglaw}) and $B_2 / \kappa$ does not tend to a constant in the high Reynolds number limit.

This discussion naturally brings us to the non-universality of measured von K\'arm\'an coefficients \cite{nagibchauhan08} which we now comment on before moving to the analysis of some of the highest Reynolds number DNS data currently available. If $E \sim u_\tau^2$ at high Reynolds numbers and the log-law therefore holds as a consequence of (\ref{eq:newmeanflow}), then, because of (\ref{eq:kappastar}), the von K\'arm\'an coefficient will have to be proportional to $B_1^2$ and inversely proportional to $C_s$, the number of stagnation points within a volume $\delta_\nu^3$ at the upper edge of the buffer layer. There is no a priori reason to expect $B_1$ and $C_s$ to be the same in turbulent channel and pipe flows, for example. There is therefore no a priori reason for the von K\'arm\'an coefficient to be the same in different such flows either.

In the case where the log-law does not hold because of the effect that inactive motions have on $E_+$ in (\ref{eq:newmeanflow}), data fitted by a log-law may yield different von K\'arm\'an coefficients both as a result of $\kappa_s = B_1^2 / C_s$ but also as a result of fitting mismatches.

In conclusion, whatever the scalings of $E_+$, one can expect measured values of the von K\'arm\'an coefficient to be non-universal as has indeed been recently reported \cite{nagibchauhan08}.
\section{High Reynolds number DNS data}
\begin{figure}[!h]
   \centering
   \includegraphics[width=8.5cm]{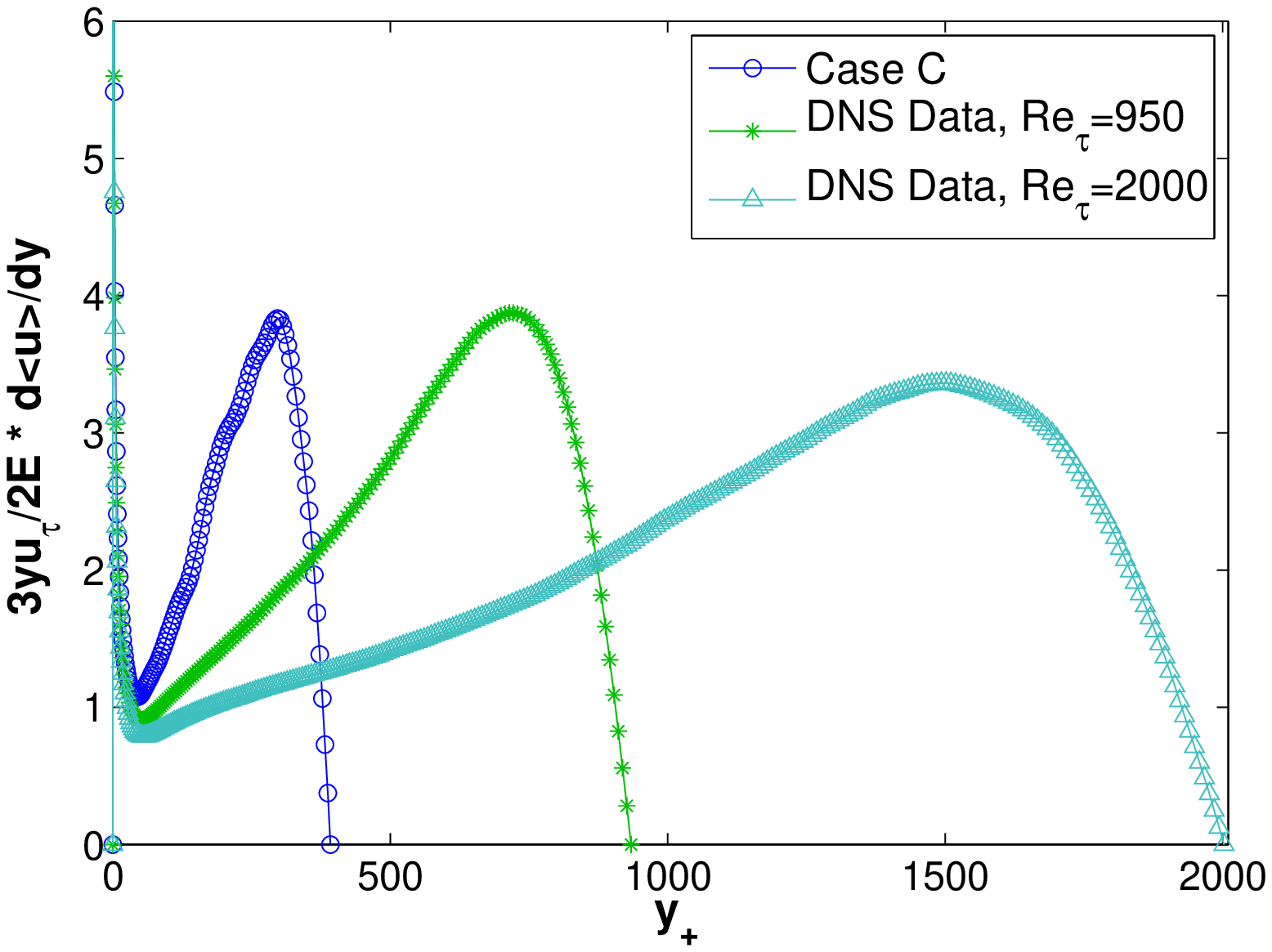}
   \includegraphics[width=8.5cm]{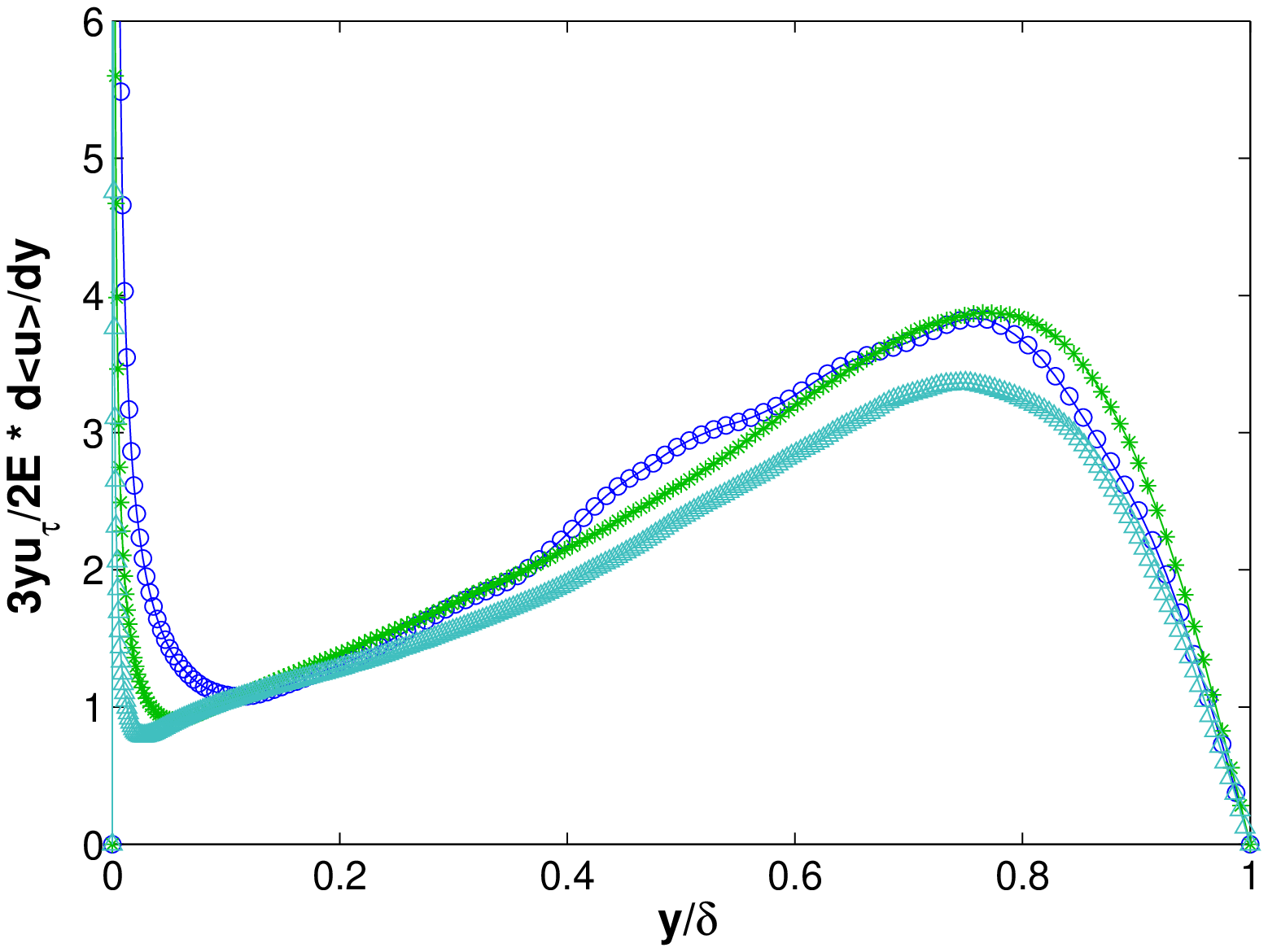}
   \caption{(Color online) $\frac{3}{2}\frac{y}{E_+ u_\tau}\frac{\dd}{\dd y} \avg u = \frac{B_2}{B_3 \kappa_s}$ as function of (a) $y_+$ and (b) $y / \delta$. DNS of turbulent channel flows without wall actuations. The $\RE_\tau =$ 950 and 2000 data are from \cite{jimenezdnsdata}}
   \label{fig:b2b3invkappastar}
\end{figure}

\begin{figure}[!h]
   \centering
   \includegraphics[width=8.5cm]{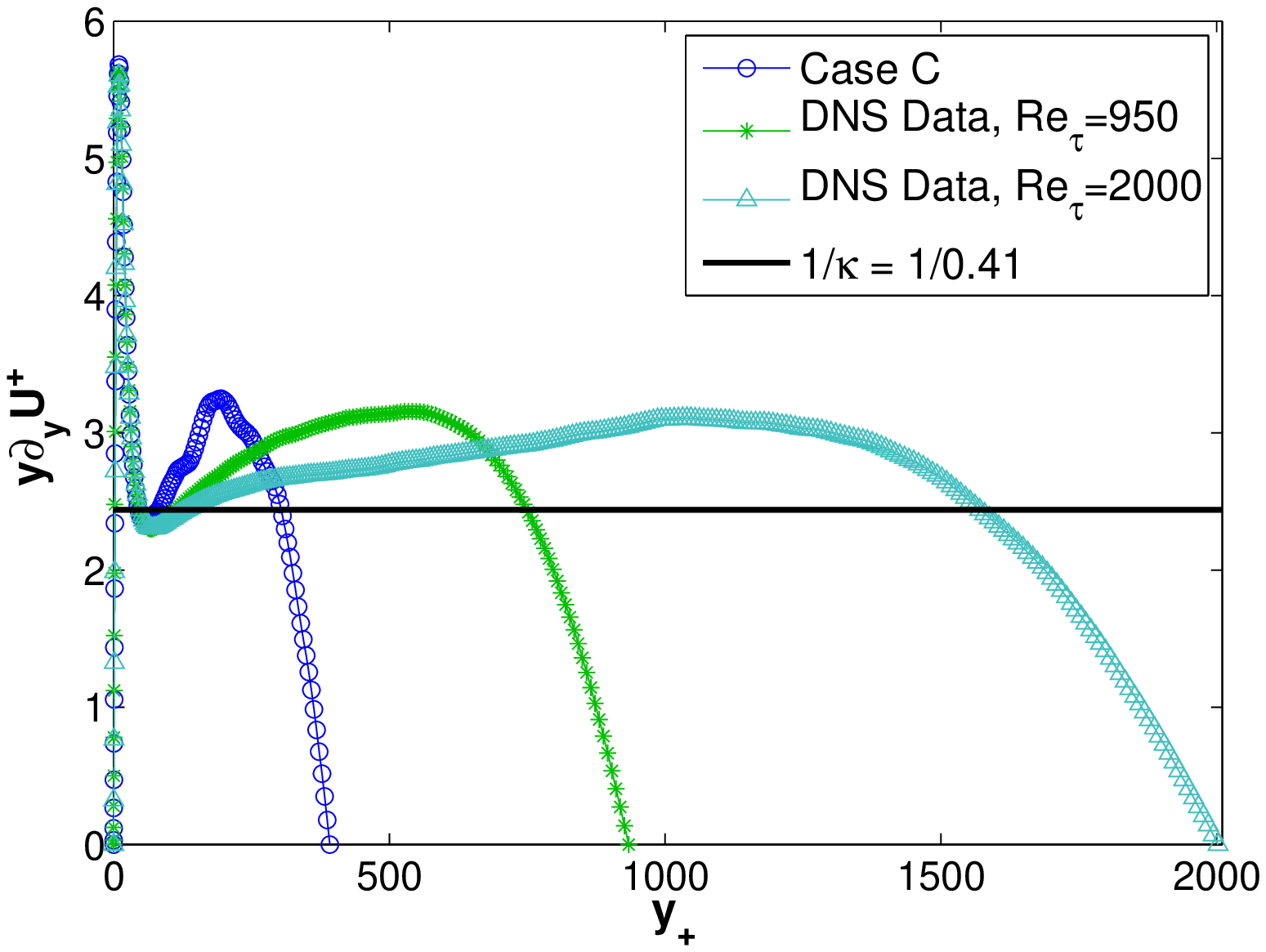}
   \includegraphics[width=8.5cm]{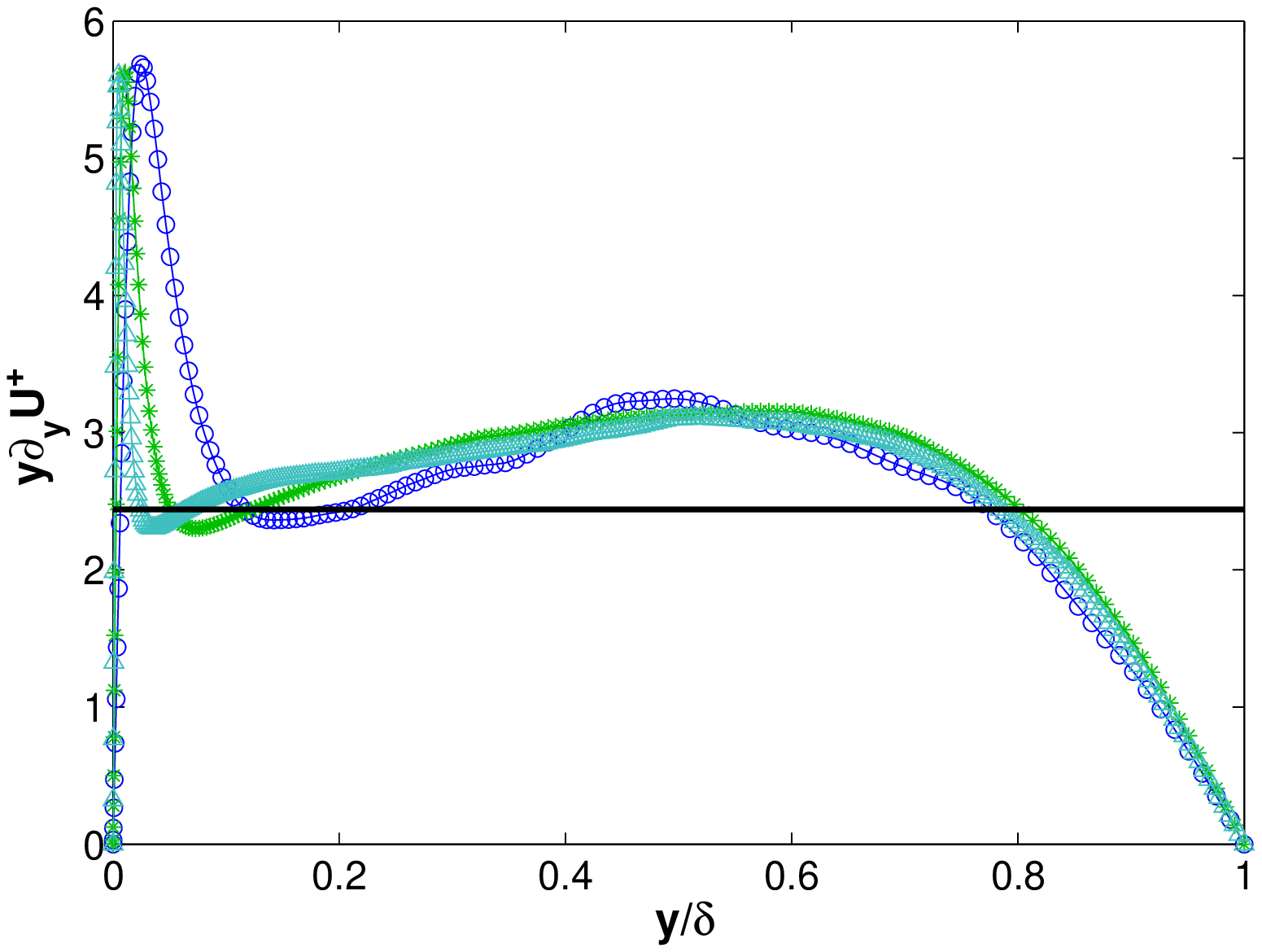}
   \caption{(Color online) $\frac{y}{u_\tau}\frac{\dd}{\dd y} \avg u = \frac{B_2}{B_3 \kappa}$ as function of (a) $y_+$ and (b) $y / \delta$. DNS of turbulent channel flows without wall actuations. The $\RE_\tau =$ 950 and 2000 data are from \cite{jimenezdnsdata}; these plots have have already been presented in \cite{jimenezdnsdata} for this data}
   \label{fig:b2b3invkappa}
\end{figure}

\begin{figure}[!h]
   \centering
   \includegraphics[width=8.5cm]{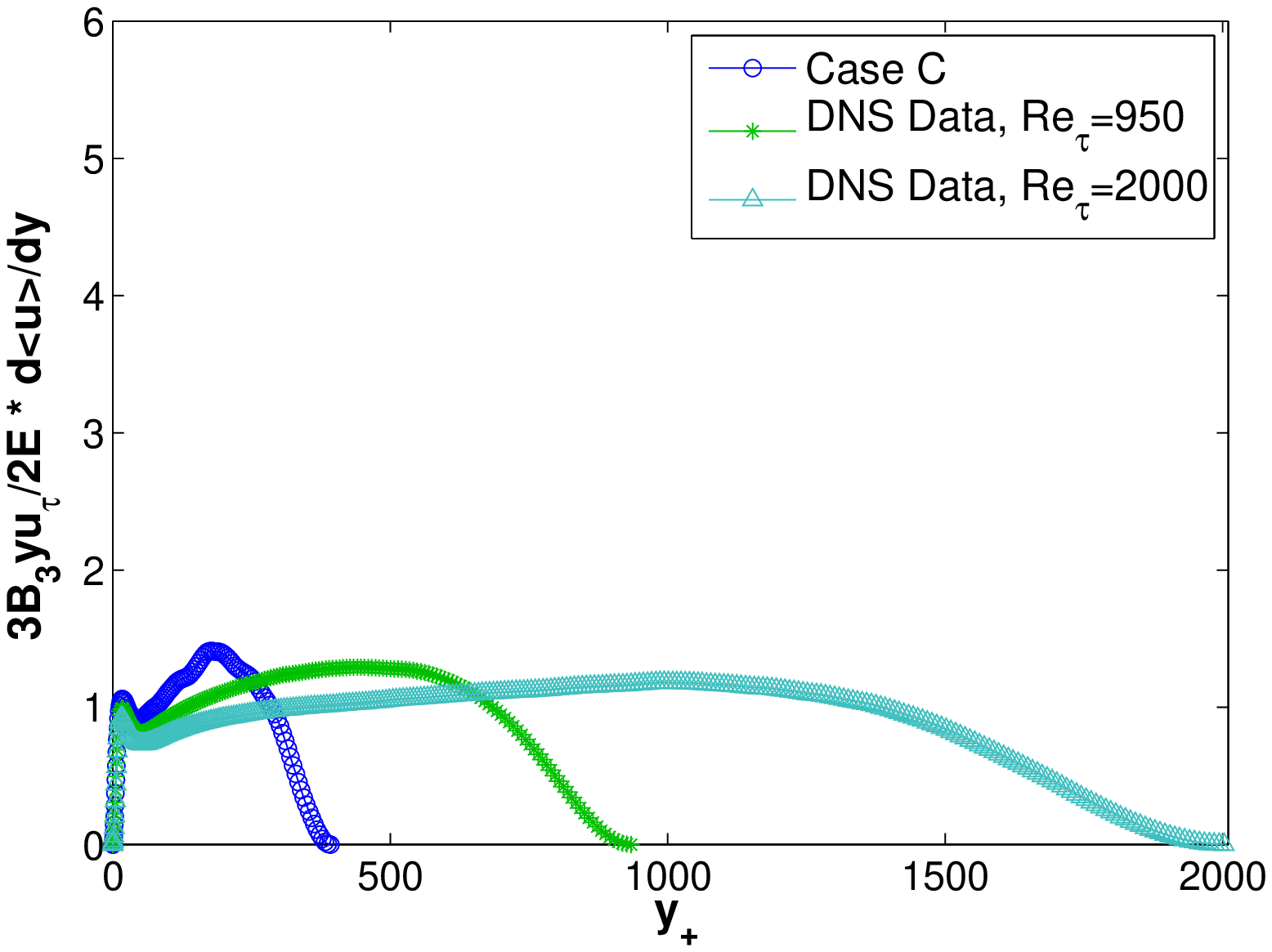}
   \includegraphics[width=8.5cm]{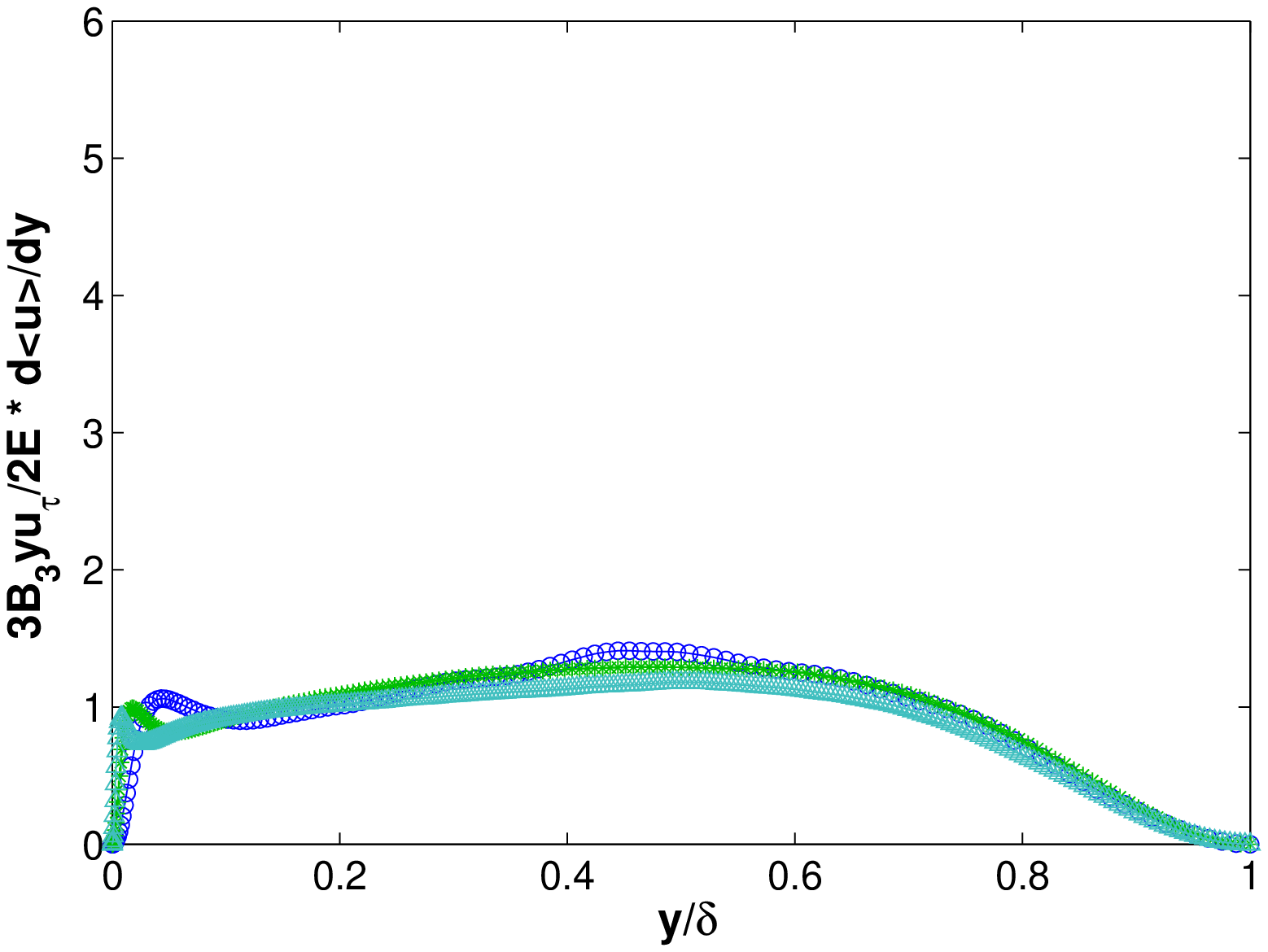}
   \caption{(Color online) $\frac{3}{2} B_3 \frac{y}{E_+ u_\tau}\frac{\dd}{\dd y} \avg u = \frac{B_2}{\kappa_s}$ as function of (a) $y_+$ and (b) $y / \delta$. DNS of turbulent channel flows without wall actuations. The $\RE_\tau =$ 950 and 2000 data are from \cite{jimenezdnsdata}}
   \label{fig:b2invkappastar}
\end{figure}

\begin{figure}[!h]
   \centering
   \includegraphics[width=8.5cm]{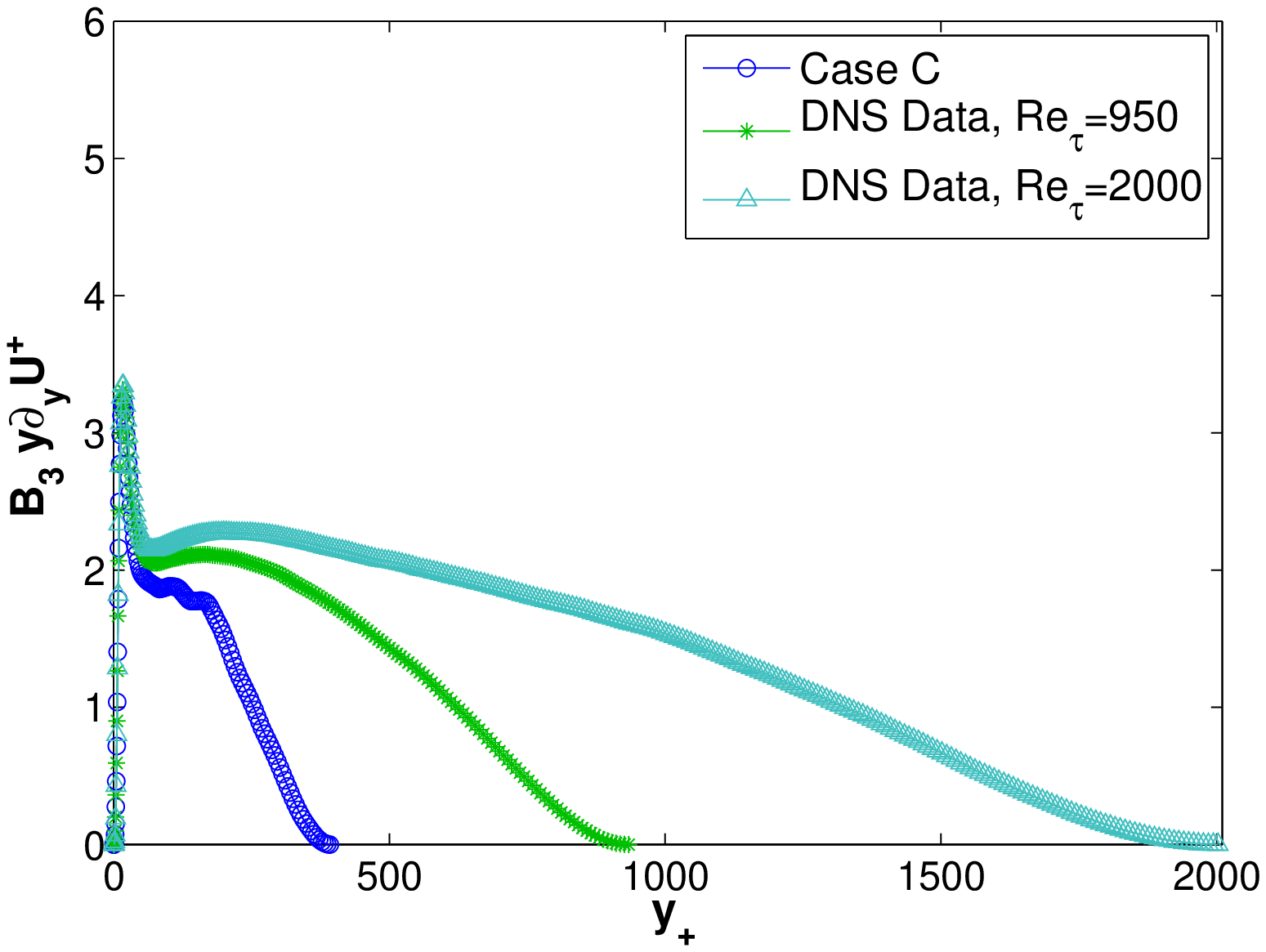}
   \includegraphics[width=8.5cm]{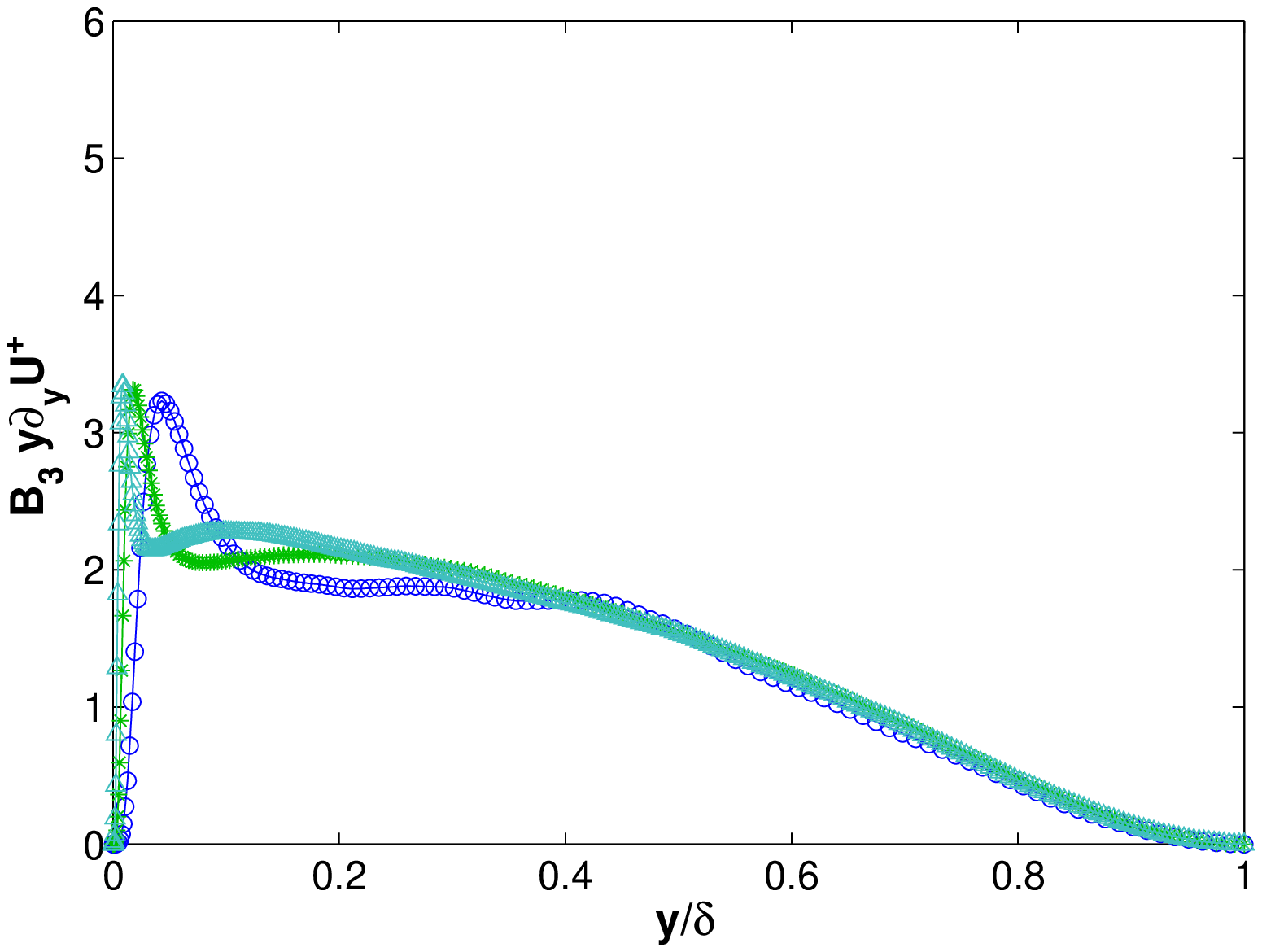}
   \caption{(Color online) $B_3\frac{y}{u_\tau}\frac{\dd}{\dd y} \avg u = \frac{B_2}{\kappa}$ as function of (a) $y_+$ and (b) $y / \delta$. DNS of turbulent channel flows without wall actuations. The $\RE_\tau =$ 950 and 2000 data are from \cite{jimenezdnsdata}}
   \label{fig:b2invkappa}
\end{figure}

We now test some of our results and conclusions on a set of data which includes the highest Reynolds number channel flow computations currently available \cite{jimenezdnsdata}, i.e. $\RE_\tau = 2000$. This set also includes data for $\RE_\tau = 950$ \cite{jimenezdnsdata} and our own highest Reynolds number DNS data ($\RE_\tau = 395$). We plot $\frac{3}{2} \frac{y}{E_+ u_\tau} \frac{\dd}{\dd y}\avg u = B_2 / (B_3 \kappa_s)$ (see Fig. \ref{fig:b2b3invkappastar}) and $\frac{y}{u_\tau} \frac{\dd}{\dd y}\avg u = B_2 / (B_3 \kappa )$ (see Fig. \ref{fig:b2b3invkappa}) as well as $\frac{3}{2} B_3 \frac{y}{E_+ u_\tau} \frac{\dd}{\dd y}\avg u = B_2 / \kappa_s$ (see Fig. \ref{fig:b2invkappastar}) and $B_3 \frac{y}{u_\tau} \frac{\dd}{\dd y}\avg u = B_2 / \kappa$ (see Fig. \ref{fig:b2invkappa}). 


\begin{figure}[!h]
   \centering
   \includegraphics[width=8.5cm]{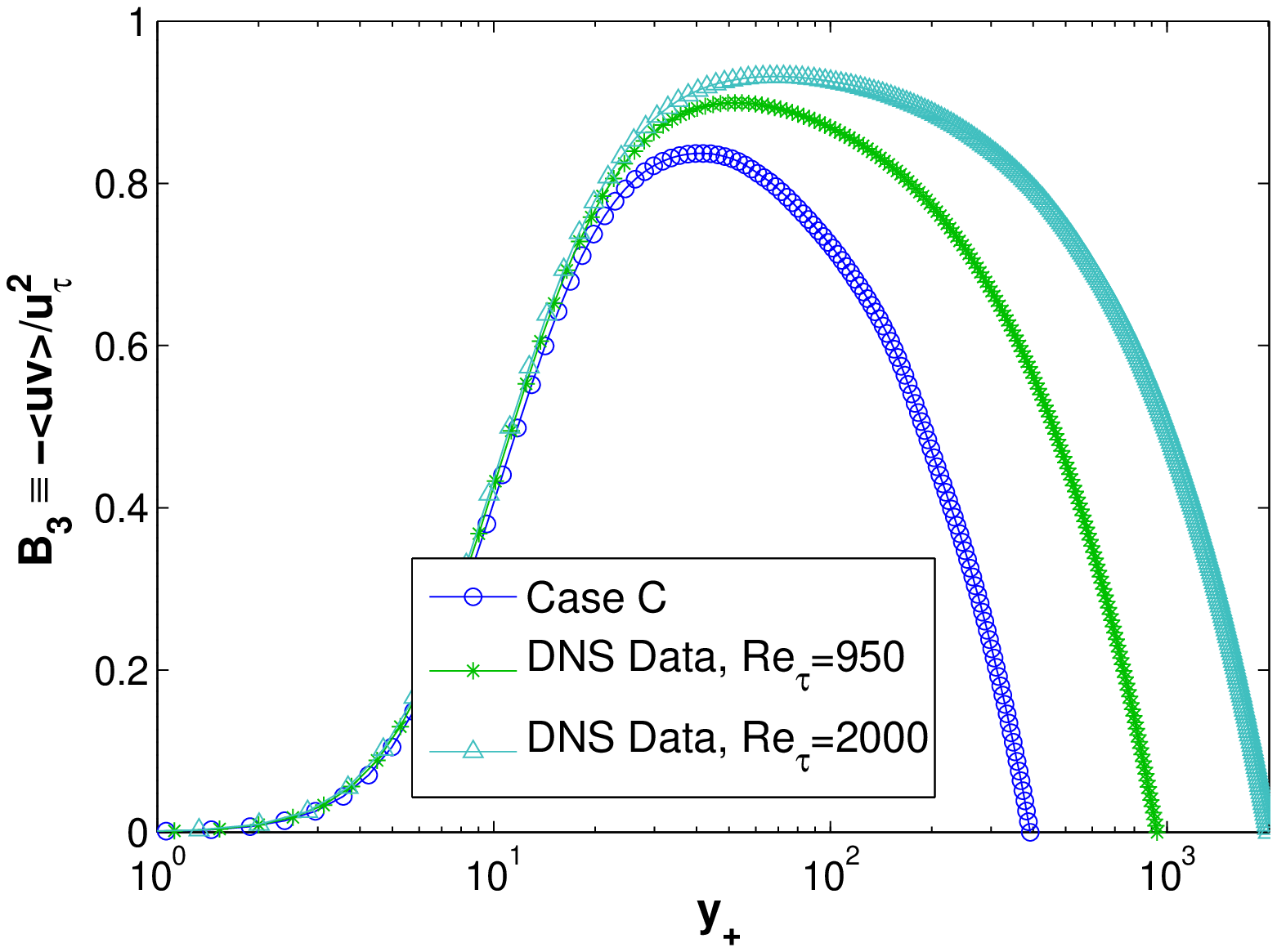}
   \includegraphics[width=8.5cm]{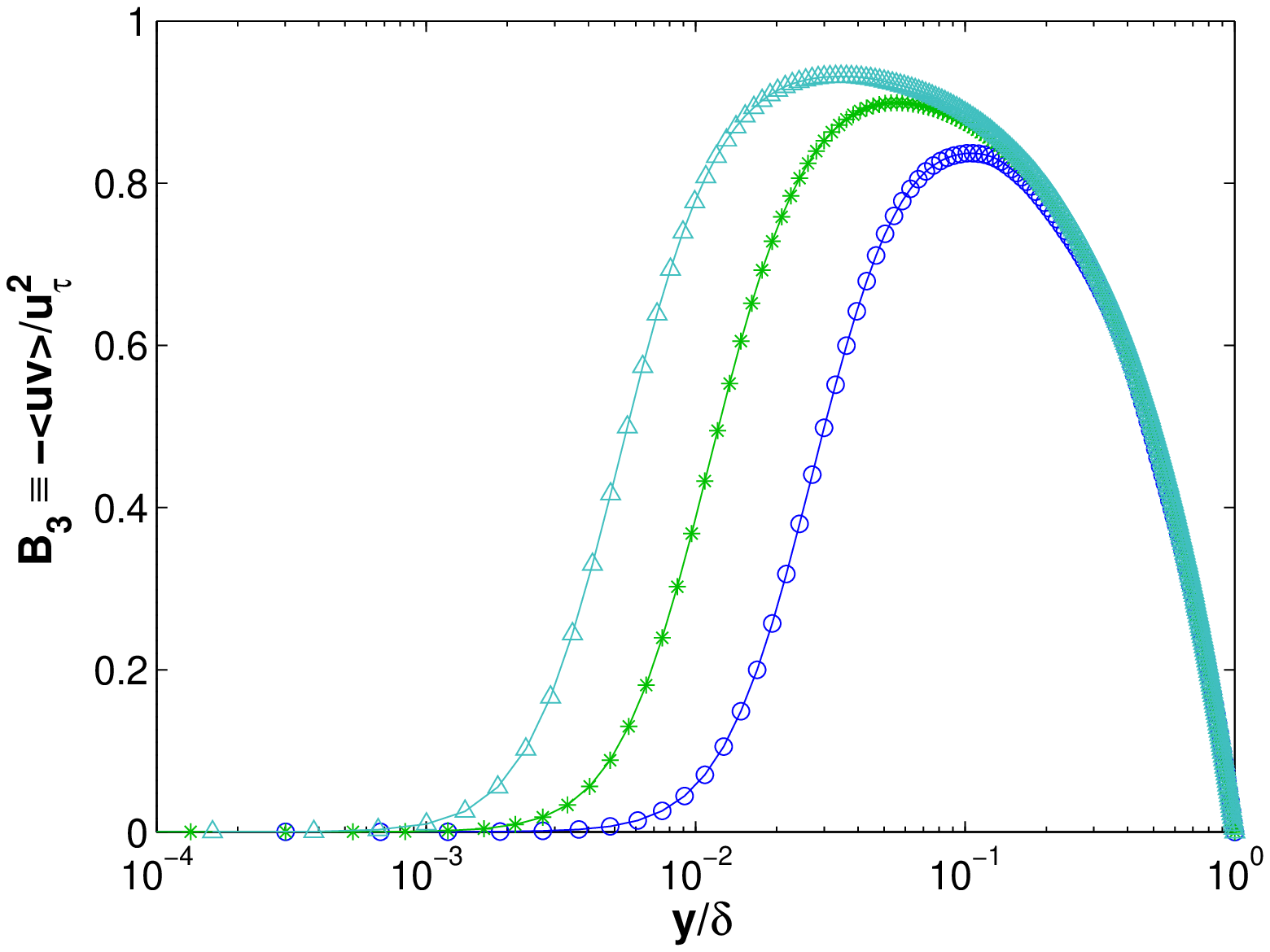}
   \caption{(Color online) $B_3 \equiv -\avg{uv} / u_\tau^2$ as function of (a) $y_+$ and (b) $y / \delta$. DNS of turbulent channel flows without wall actuations. The $\RE_\tau =$ 950 and 2000 data are from \cite{jimenezdnsdata}}
   \label{fig:b3}
\end{figure}

These high Reynolds number results support and extend the claims made in the previous section: $B_2 / \kappa_s$ appears to have the least departures from constancy in the intermediate range, better that $B_2 / (B_3 \kappa)$ which is however better than $B_2 / (B_3 \kappa_s)$. The variations of $B_2 / \kappa$ are offset by those of $B_3$ (see Fig. \ref{fig:b3}) which explains why $B_2 / (B_3 \kappa)$ looks better than $B_2 / (B_3 \kappa_s)$.
The situation remains therefore identical to the one we encountered with the lower Reynolds number simulations in the previous section. The Reynolds number needs to be very much larger than 2000 for $B_3$ to come close to its asymptotic value 1 in an intermediate layer, as already shown by experimental measurements spanning an ever wider Reynolds number range in \cite{nagibchauhan08}.

\begin{figure}[!h]
   \centering
\includegraphics[width=8.5cm]{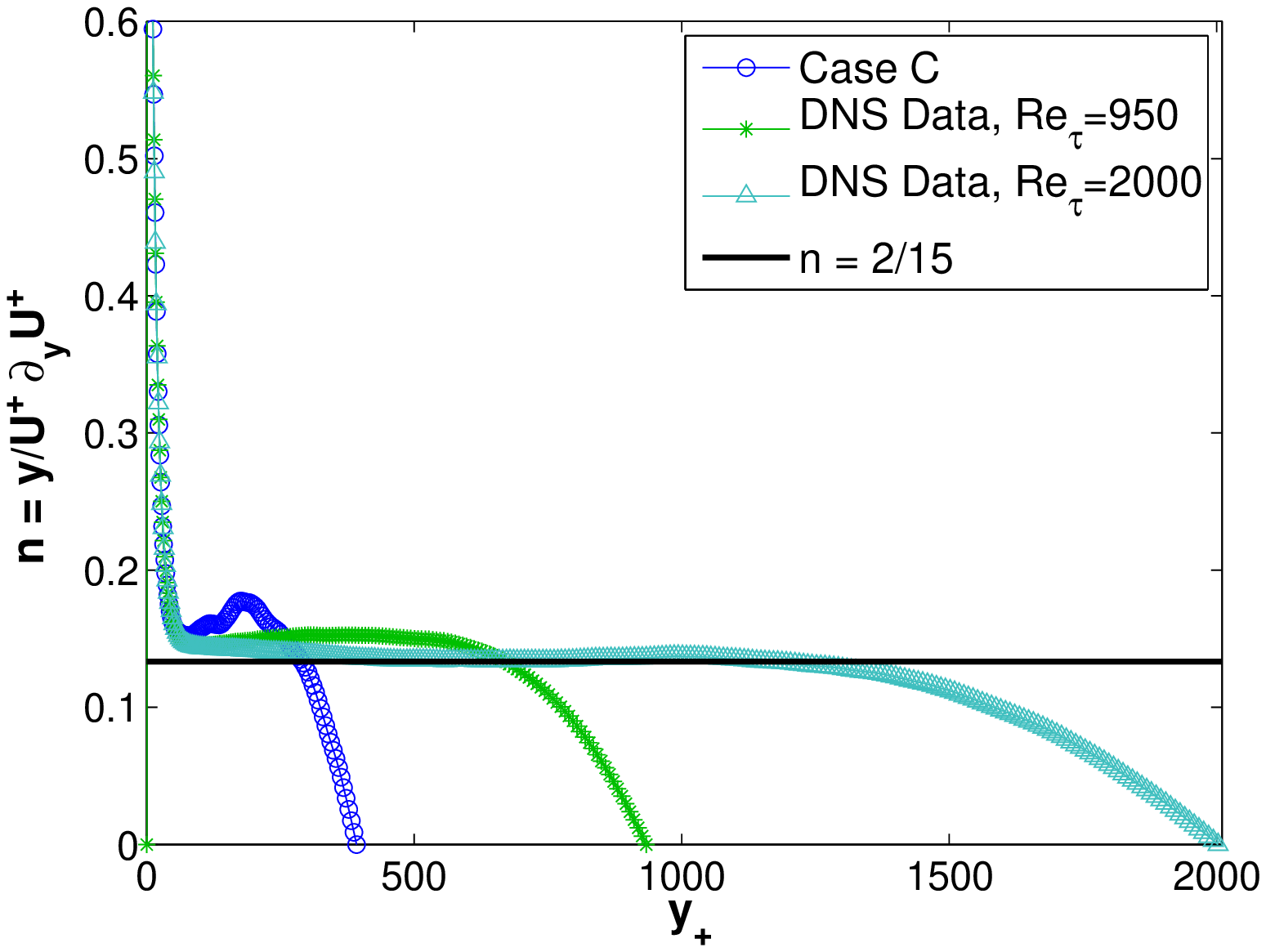}
   \caption{(Color online) Power law mean velocity profile: $n = \frac{y}{U_+} \frac{\dd}{\dd y}U_+$ plotted against $y_+$. DNS of turbulent channel flows without wall actuations. The $\RE_\tau =$ 950 and 2000 data are from \cite{jimenezdnsdata}}
   \label{fig:powerlaw}
\end{figure}

\begin{figure}[!h]
   \centering
\includegraphics[width=8.5cm]{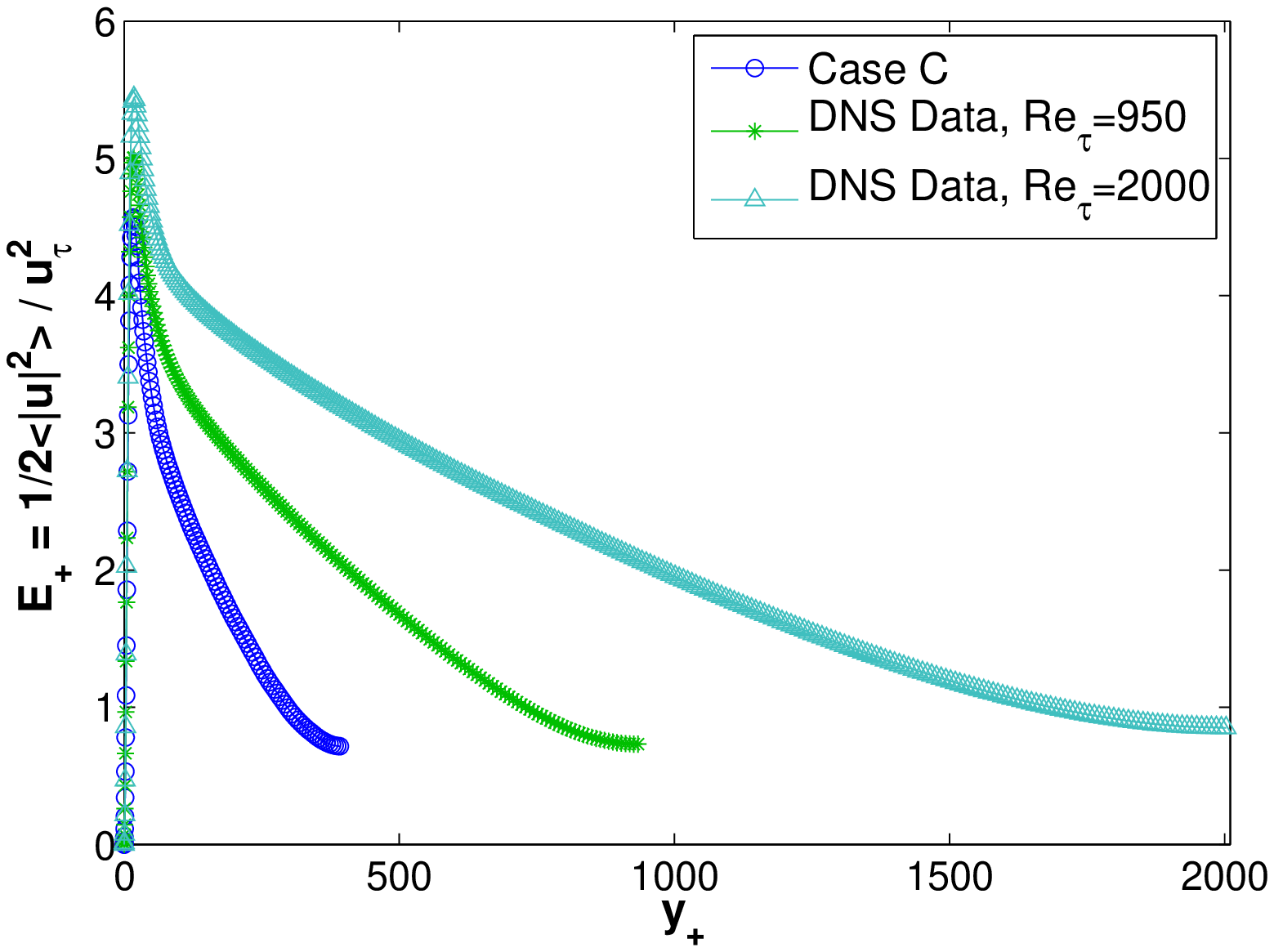}
\includegraphics[width=8.5cm]{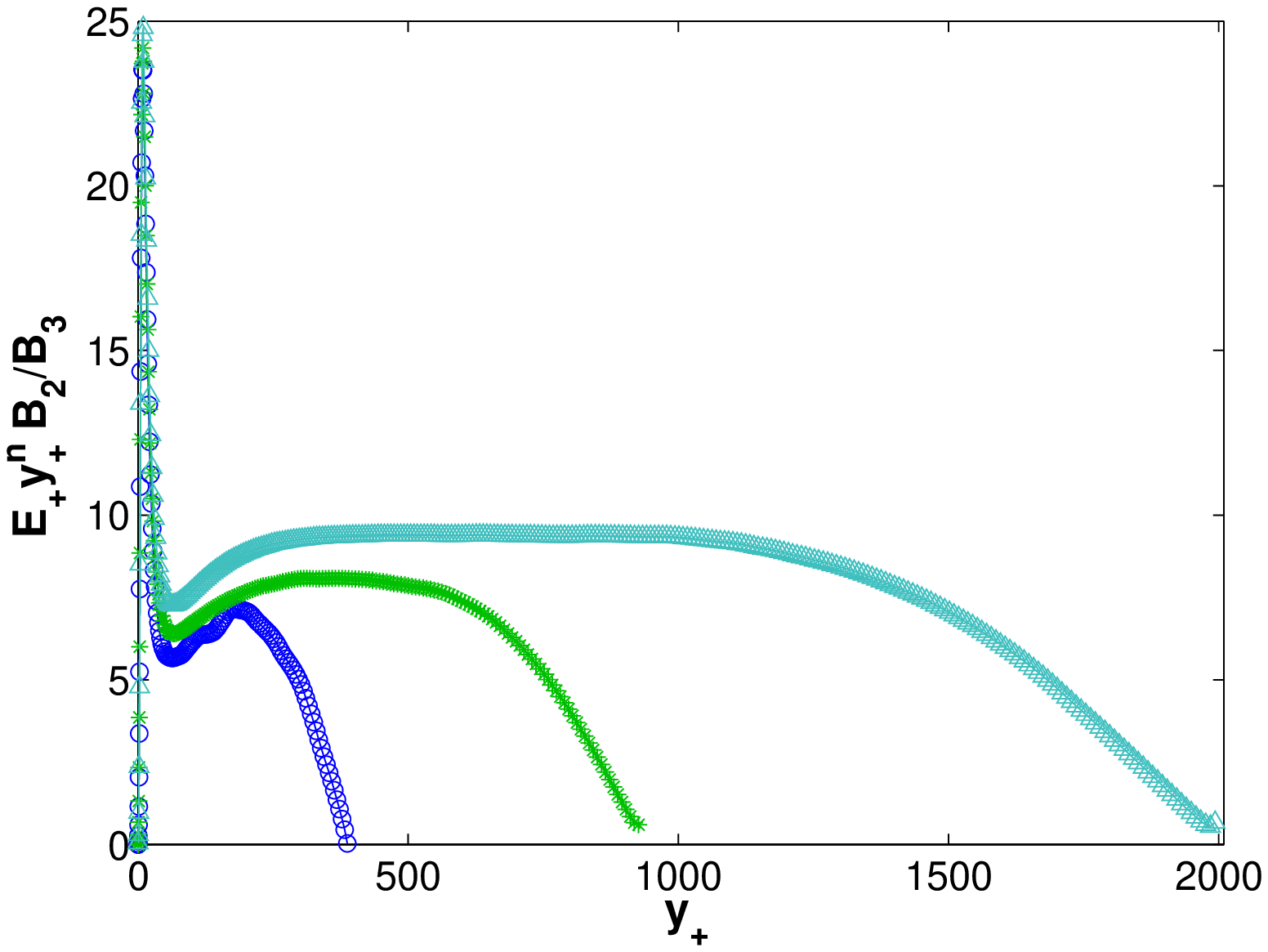}
   \caption{(Color online) (a) $E_+$ and (b) $E_+ y_+^n \frac{B_2}{B_3}$ with $n=\frac{2}{15}$ versus $y_+$ for DNS of turbulent channel flows without wall actuations. The $\RE_\tau =$ 950 and 2000 are from \cite{jimenezdnsdata}}
   \label{fig:Eplus}
\end{figure}

Alternative forms for the mean flow profile at high Reynolds numbers have been proposed in the literature and we test in Fig. \ref{fig:powerlaw} the suggestion of a power-law form \cite{barenblattetal97,george07}. The high Reynolds number data we are using here appear to give significant support to such a power-law form with power exponent $n \equiv \frac{y}{U_+} \frac{\dd}{\dd y} U_+ \simeq 2/15$, i.e. $\frac{\dd}{\dd y_+}U_+ \sim y_+^{-(1 + 2/15)}$ in the intermediate layer.
On the basis of (\ref{eq:newmeanflow}), this result suggests that $E_+$ has a power-law dependence on $y_+$ in that same layer. Indeed, combining (\ref{eq:newmeanflow}) in its finite Reynolds number form, i.e. $\frac{\dd}{\dd y}\avg u \simeq \frac{2}{3} \frac{B_2}{B_3} E_+ \frac{u_\tau}{\kappa_s y}$, with $\frac{\dd}{\dd y_+}U_+ \simeq \frac{B_4}{\kappa_s} y_+^{-(1 + 2/15)}$ yields $E_+ y_+^{2/15} \frac{B_2}{B_3} \simeq \frac{3}{2} B_4$, i.e. a constant value of $E_+ y_+^{2/15} B_2 / B_3$ in the equilibrium layer if $B_4$ is constant in that layer. Figure \ref{fig:Eplus} supports this conclusion albeit with a constant value of $E_+ y_+^{2/15} B_2 / B_3$ which appears to increase slowly with Reynolds number. This Reynolds number dependence may be intrinsic to $E_+$ resulting, perhaps, from Townsend's inactive eddy hypothesis.
\section{Conclusion: a new asymptotic approach}
Our DNS suggest that $B_1 = \lambda / \ell_s$ and $C_s = n_s \delta_\nu^3 y_+$ are approximately constants in the region $\delta_\nu \ll y \lesssim \delta$. Our DNS demonstrate these well-defined approximate constancies for $\RE_\tau$ as low as a few hundred. These constancies imply that, in the region $\delta_\nu \ll y \lesssim \delta$, the eddy turnover time $\tau$ equals $\tfrac{3}{2}\kappa_sy / u_\tau$ with $\kappa_s = B_1^2 / C_s$. Assuming the constancies of $B_1$ and $C_s$ to be early manifestations of a high Reynolds number behaviour, i.e. that $B_1$ and $C_s$ and therefore $\kappa_s$ are independent of $y$ in $\delta_\nu \ll y \lesssim \delta$ as $\RE_\tau \to \infty$, it then follows that $\tfrac{\dd \avg{u}}{\dd y} = \frac{2}{3} E_+ \tfrac{u_\tau}{\kappa_s y}$ in the equilibrium region $\delta_\nu \ll y \ll \delta$ where production may be expected to balance dissipation and $-\avg{uv} \simeq u_\tau^2$. The asymptotic equality $-\avg{uv} \simeq u_\tau^2$ is mathematically supported only for turbulent channel/pipe flows.

The classical intermediate asymptotics approach, which assumes no dependence of the mean velocity gradient on $\nu$ and $\delta$ where $\delta_\nu \ll y \ll \delta$, does not consider the effect that Townsend's inactive motions \cite{townsend76} may or may not have on $\tfrac{\dd}{\dd y}\avg u$. However, if a new intermediate asymptotic approach is taken where the assumption of no dependence on $\nu$ and $\delta$ is applied to $\tau \equiv E / \epsilon$ instead of $\tfrac{\dd}{\dd y}\avg u$, we are then led to $\tau \sim y / u_\tau$, and the small effects of inactive motions on $\tfrac{\dd}{\dd y}\avg u$ may not be neglected because $\tau \simeq \frac{3}{2} \kappa_s y / u_\tau$, $-\avg{uv} \simeq u_\tau^2$ and $-\avg{uv}\tfrac{\dd}{\dd y}\avg{u} \simeq \epsilon$ yield (\ref{eq:newmeanflow}) which explicitly contains $E_+$. If as a result of inactive motions, $E$ does not scale as $u_\tau^2$ then this revised intermediate asymptotic approach will not predict a log law for the mean profile even though a stagnation point von K\'arm\'an coefficient $\kappa_s$ exists and is well defined within the approach. The mean flow prediction of this approach is instead controlled by the intermediate asymptotic dependence of $E_+$ on $y_+$ and $\RE_\tau$. If this dependence on $y_+$ is a power law $E_+ \sim y_+^{-n}$ in the intermediate range $\delta_\nu \ll y \ll \delta$, then the mean flow profile will also be a power law, i.e. $\frac{\dd}{\dd y_+} U_+ \sim y_+^{-1-n}$, in that intermediate layer.

DNS of turbulent channel flow with the highest values of $\RE_\tau$ currently available \cite{jimenezdnsdata} suggest the same $n = 2/15$ in both $E_+ \sim y_+^{-n}$ and $\frac{\dd}{\dd y_+} U_+ \sim y_+^{-1-n}$ in support of our procedure and formula (\ref{eq:newmeanflow}). However, we should caution against extrapolating this value of the exponent $n$ to higher values of $Re_{\tau}$, in particular in the laboratory where the boundary conditions are in fact different from the DNS which is periodic in $x$ and $z$. We stress that the main point of value here for us is the support that these exceptional DNS \cite{jimenezdnsdata} bring to our approach and in particular to our formula (\ref{eq:newmeanflow}).

Note also that the stagnation point von K\'arm\'an coefficient is defined by $\epsilon \simeq \frac{2}{3}E_+ \frac{u_\tau^3}{\kappa_s y}$ in the range $\delta_\nu \ll y \lesssim \delta$ irrespective of whether the mean flow profile is a log law or a power law. Power law profiles of $E_+$ and $\frac{\dd}{\dd y_+}U_+$ in the intermediate layer $\delta_\nu \ll y \ll \delta$, simply force $\epsilon \simeq \frac{2}{3} E_+ \frac{u_\tau^3}{\kappa_s y}$ to imply that the classical relation $\epsilon \simeq \frac{u_\tau^3}{\kappa y}$ does not hold in that layer.

This new intermediate asymptotic approach is supported by our DNS observations that (\ref{eq:ricethm}) and (\ref{eq:-1law}) are valid in the region $\delta_\nu \ll y \lesssim \delta$ because (\ref{eq:ricethm}) and (\ref{eq:-1law}) imply $\tau \simeq \frac{3}{2} \kappa_s y / u_\tau$ with $\kappa_s = B_1^2 / C_s$. The relation $\kappa_s = B_1^2 / C_s$ offers a link between the underlying flow structure, described in terms of stagnation points, and the dissipation/cascade statistics of the turbulence. The universality in terms of both $\RE_\tau$ and flow-type dependencies of $\kappa_s$ becomes a question concerning the universality of the stagnation point structure of the turbulent fluctuations. To what extent does it depend on boundary and wall forcing conditions? Is it the same in turbulent channel and turbulent pipe flows? Is it the same in DNS of such flows where periodic boundary conditions are used and in laboratory realisations of such flows where boundary conditions are clearly not periodic? These are questions which must be left for future investigation, but our approach makes them fully legitimate as there is no reason to expect the stagnation point structure of turbulent fluctuations to be exactly the same in all these cases.

The implications of this new approch for the mean flow profile in turbulent channel/pipe flows come by invoking a local balance between production and dissipation as well as $-\avg{uv} \simeq u_\tau^2$ in the intermediate range $\delta_\nu \ll y \ll \delta$ as $\RE_\tau \to \infty$. A direct test against data of $\frac{\dd}{\dd y_+} U_+ \simeq \frac{2}{3} \frac{E_+}{\kappa_s y_+}$ in that same range and limit cannot be expected to be successful if $\RE_\tau$ is not large enough for $-\avg{uv}$ to equal $u_\tau^2$ over the range $\delta_\nu \ll y \ll \delta$. As clearly shown by various experimental and numerical data, this equality is well beyond the highest Reynolds numbers currently available both numerically and in the laboratory. The siginificant finite Reynolds number deviations from $-\avg{uv} \simeq u_\tau^2$ appears to compensate the deviations from a log-law and from the local production-dissipation balance with the result that plots of $y \frac{\dd}{\dd y} U_+$ have a less varying appearance than plots of $\frac{3}{2} \frac{y}{E_+} \frac{\dd}{\dd y} U_+$. At face value this could be misinterpreted as better support for the log-law $\frac{\dd}{\dd y_+} U_+ \simeq \frac{1}{\kappa y_+}$ than for $\frac{\dd}{\dd y_+} U_+ \simeq \frac{2}{3} \frac{E_+}{\kappa_s y_+}$. However, plots of $B_2 / \kappa_s$ look significantly better than plots of $B_2 / \kappa$ thus demonstrating that $\frac{3}{2} \frac{y}{E_+} \frac{\dd}{\dd y} U_+ = B_2 / (B_3 \kappa_s)$ looks worse than $y \frac{\dd}{\dd y} U_+ = B_2 / (B_3 \kappa)$ only because $B_3 = -\avg{uv} / u_\tau^2$ is so significantly non-constant (see Fig. \ref{fig:b3}).

As a final note, future investigations should attempt to uncover the small-scale intermittency corrections to our scalings by determining the weak dependencies that $B_1$ may have on $\RE_\tau$ and $y_+$ as a result of small-scale intermittency. These dependencies will cause dependencies of $\kappa_s$ on $\RE_\tau$ and $y_+$. It is remarkable that small-scale intermittency may have an impact, even if small, on the scalings of mean flow profiles. These Reynolds number parts of their scalings should be distinguished from those that Townsend's attached eddies may be contributing via the scaling of $E_+$ on $\RE_\tau$.
\begin{acknowledgments}
We are grateful to Dr. Sylvain Laizet for providing the Navier-Stokes solver and to Halliburton for the financial support. We also thank S. Hojas \& J. Jimenez for making their data available on the web.
\end{acknowledgments}
\appendix*
\section{Numerical method for the computation of stagnation points}
\label{app:stgnptsmethod}
In this study of turbulent channel flow, we focus on stagnation points of the fluctuating velocity field, i.e.
\begin{equation}
   \label{eq:zerovelpts}
   \bm u'(\bm x, t) \equiv \bm u(\bm x, t) - \avg{\bm u} = 0
\end{equation}
where $\avg.$ here, denotes an average in space over the homogeneous directions $x$ and $z$ at a particular time. These zero-velocity points are Galilean invariant and result from the intersections of the three random surfaces $u'(\bm x, t) = 0$, $v'(\bm x, t) = 0$, $w'(\bm x, t) = 0$. Intersections of two random surfaces gives lines and the intersections of these lines with a third random surface gives points.

A root finding method is required to obtain where in space our random function $\bm u'(\bm x, t)$ is locally zero. Here we use the iterative Newton-Raphson method,
\begin{equation}
   \label{eq:newton}
   \bm x_{new} = \bm x_{old} + \delta \bm x ~\text{with}~ \lt[ \grad \bm u'
\rt]_{\mathcal L} \delta \bm x = - \bm u'_{\mathcal L}
\end{equation}
where $\grad \bm u'(\bm x, t) \equiv \grad \bm u(\bm x, t) - \avg{\grad
\bm u}$ and the subscript $\mathcal L$ stands for an interpolated quantity. The interpolation of the velocity and its gradient was done using fourth-order Lagrangian interpolation \cite{nr}. The particular choice of the interpolation was chosen based on robustness, accuracy and computational efficiency. The $3 \times 3$ linear system of equations was solved for $\delta \bm x$ simply using the Cramer's rule
\begin{widetext}
\begin{equation}
   \label{eq:cramer}
   \delta \bm x = - \frac
   { \lt[ \text{det}(\bm u'_{\mathcal L},
                             \tfrac{\pd\bm u'}{\pd y}\Lint,
                             \tfrac{\pd\bm u'}{\pd z}\Lint),
          \text{det}(\tfrac{\pd\bm u'}{\pd x}\Lint,
                             \bm u'_{\mathcal L},
                             \tfrac{\pd\bm u'}{\pd z}\Lint),
          \text{det}(\tfrac{\pd\bm u'}{\pd x}\Lint,
                             \tfrac{\pd\bm u'}{\pd y}\Lint,
                             \bm u'_{\mathcal L}) \rt] }
   { \text{det}(\tfrac{\pd\bm u'}{\pd x}\Lint,
                         \tfrac{\pd\bm u'}{\pd y}\Lint,
                         \tfrac{\pd\bm u'}{\pd z}\Lint) }
\end{equation}
\end{widetext}
assuming that $\text{det}( \tfrac{\pd\bm u'}{\pd x}\Lint,\tfrac{\pd\bm
u'}{\pd y}\Lint,\tfrac{\pd\bm u'}{\pd z}\Lint ) \neq 0$.

The Newton-Raphson method gives a very rapid local convergence to a root if the initial guess is sufficiently good. So to have a chance at good starting points we took them midway between two consecutive grid points throughout our computational domain. It is well known that different initial guesses can converge to the same solution, due to the unpredictable global convergence properties of this iterative method. To avoid this issue we try to bound the root finding no further than the neighbouring computational cells and we also ensure that no more than one stagnation point exists in a cell, which is what should be expected from a smooth velocity field of a well resolved DNS. 

Our method was also compared to an algorithm proposed in \cite{schmelcherdiakonos98} for the detection of unstable periodic orbits in chaotic dynamical systems, which has good global convergence due to its attracting nature. This method is based on a universal set of linear transformations, which transform unstable fixed points to stable ones whilst mainting their positions. However, this method can be expensive in more than two dimensions and this is the reason why we choose here Newton-Raphson, which is as accurate.

The number of zero-velocity points $N_s$ are computed within thin slabs of dimensions $L_x \times \delta_y \times L_z$, with $\delta_y \sim \delta_\nu$, parallel to the channel's wall. Time averages of $N_s$ were taken by repeating the same procedure for several time instances.
\newpage 
\bibliography{references}

\begin{thebibliography}{27}
\expandafter\ifx\csname natexlab\endcsname\relax\def\natexlab#1{#1}\fi
\expandafter\ifx\csname bibnamefont\endcsname\relax
  \def\bibnamefont#1{#1}\fi
\expandafter\ifx\csname bibfnamefont\endcsname\relax
  \def\bibfnamefont#1{#1}\fi
\expandafter\ifx\csname citenamefont\endcsname\relax
  \def\citenamefont#1{#1}\fi
\expandafter\ifx\csname url\endcsname\relax
  \def\url#1{\texttt{#1}}\fi
\expandafter\ifx\csname urlprefix\endcsname\relax\def\urlprefix{URL }\fi
\providecommand{\bibinfo}[2]{#2}
\providecommand{\eprint}[2][]{\url{#2}}

\bibitem[{\citenamefont{Zanoun et~al.}(2003)\citenamefont{Zanoun, Durst, and
  Nagib}}]{zanounetal03}
\bibinfo{author}{\bibfnamefont{E.}~\bibnamefont{Zanoun}},
  \bibinfo{author}{\bibfnamefont{F.}~\bibnamefont{Durst}}, \bibnamefont{and}
  \bibinfo{author}{\bibfnamefont{H.}~\bibnamefont{Nagib}},
  \bibinfo{journal}{Phys. Fluids} \textbf{\bibinfo{volume}{15}},
  \bibinfo{pages}{3079} (\bibinfo{year}{2003}).

\bibitem[{\citenamefont{Zagarola and Smits}(1998)}]{zagarolasmits98}
\bibinfo{author}{\bibfnamefont{M.~V.} \bibnamefont{Zagarola}} \bibnamefont{and}
  \bibinfo{author}{\bibfnamefont{A.~J.} \bibnamefont{Smits}},
  \bibinfo{journal}{Journal of Fluid Mechanics} \textbf{\bibinfo{volume}{373}},
  \bibinfo{pages}{33} (\bibinfo{year}{1998}).

\bibitem[{\citenamefont{Nagib and Chauhan}(2008)}]{nagibchauhan08}
\bibinfo{author}{\bibfnamefont{H.~M.} \bibnamefont{Nagib}} \bibnamefont{and}
  \bibinfo{author}{\bibfnamefont{K.~A.} \bibnamefont{Chauhan}},
  \bibinfo{journal}{Phys. Fluids} \textbf{\bibinfo{volume}{20}},
  \bibinfo{eid}{101518} (\bibinfo{year}{2008}).

\bibitem[{\citenamefont{Pope}(2000)}]{pope00}
\bibinfo{author}{\bibfnamefont{S.~B.} \bibnamefont{Pope}},
  \emph{\bibinfo{title}{{Turbulent Flows}}} (\bibinfo{publisher}{Cambridge
  University Press}, \bibinfo{year}{2000}).

\bibitem[{\citenamefont{L'vov et~al.}(2008)\citenamefont{L'vov, Procaccia, and
  Rudenko}}]{lvovetal08}
\bibinfo{author}{\bibfnamefont{V.~S.} \bibnamefont{L'vov}},
  \bibinfo{author}{\bibfnamefont{I.}~\bibnamefont{Procaccia}},
  \bibnamefont{and} \bibinfo{author}{\bibfnamefont{O.}~\bibnamefont{Rudenko}},
  \bibinfo{journal}{Phys. Rev. Lett.} \textbf{\bibinfo{volume}{100}},
  \bibinfo{eid}{054504} (\bibinfo{year}{2008}).

\bibitem[{\citenamefont{Mazellier and Vassilicos}(2008)}]{mv08}
\bibinfo{author}{\bibfnamefont{N.}~\bibnamefont{Mazellier}} \bibnamefont{and}
  \bibinfo{author}{\bibfnamefont{J.~C.} \bibnamefont{Vassilicos}},
  \bibinfo{journal}{Phys. Fluids} \textbf{\bibinfo{volume}{20}},
  \bibinfo{eid}{015101} (\bibinfo{year}{2008}).

\bibitem[{\citenamefont{Goto and Vassilicos}(2009)}]{gv09}
\bibinfo{author}{\bibfnamefont{S.}~\bibnamefont{Goto}} \bibnamefont{and}
  \bibinfo{author}{\bibfnamefont{J.~C.} \bibnamefont{Vassilicos}},
  \bibinfo{journal}{Phys. Fluids} \textbf{\bibinfo{volume}{21}},
  \bibinfo{eid}{035104} (\bibinfo{year}{2009}).

\bibitem[{\citenamefont{D\'avila and Vassilicos}(2003)}]{dv03}
\bibinfo{author}{\bibfnamefont{J.}~\bibnamefont{D\'avila}} \bibnamefont{and}
  \bibinfo{author}{\bibfnamefont{J.~C.} \bibnamefont{Vassilicos}},
  \bibinfo{journal}{Phys. Rev. Lett.} \textbf{\bibinfo{volume}{91}},
  \bibinfo{pages}{144501} (\bibinfo{year}{2003}).

\bibitem[{\citenamefont{Salazar and Collins}(2009)}]{salazarcollins09}
\bibinfo{author}{\bibfnamefont{J.~P.} \bibnamefont{Salazar}} \bibnamefont{and}
  \bibinfo{author}{\bibfnamefont{L.~R.} \bibnamefont{Collins}},
  \bibinfo{journal}{Ann. Rev. Fluid Mech.} \textbf{\bibinfo{volume}{41}},
  \bibinfo{pages}{405} (\bibinfo{year}{2009}).

\bibitem[{\citenamefont{Hoyas and Jimenez}(2006)}]{jimenezdnsdata}
\bibinfo{author}{\bibfnamefont{S.}~\bibnamefont{Hoyas}} \bibnamefont{and}
  \bibinfo{author}{\bibfnamefont{J.}~\bibnamefont{Jimenez}},
  \bibinfo{journal}{Phys. Fluids} \textbf{\bibinfo{volume}{18}},
  \bibinfo{pages}{011702} (\bibinfo{year}{2006}),
  \urlprefix\url{http://torroja.dmt.upm.es/ftp/channels/data/}.

\bibitem[{\citenamefont{Laizet and Lamballais}(2009)}]{laizetlamballais09}
\bibinfo{author}{\bibfnamefont{S.}~\bibnamefont{Laizet}} \bibnamefont{and}
  \bibinfo{author}{\bibfnamefont{E.}~\bibnamefont{Lamballais}},
  \bibinfo{journal}{J. Comp. Phys.} \textbf{\bibinfo{volume}{228}},
  \bibinfo{pages}{5989 } (\bibinfo{year}{2009}).

\bibitem[{\citenamefont{Cain et~al.}(1984)\citenamefont{Cain, Ferziger, and
  Reynolds}}]{cainetal84}
\bibinfo{author}{\bibfnamefont{A.}~\bibnamefont{Cain}},
  \bibinfo{author}{\bibfnamefont{J.}~\bibnamefont{Ferziger}}, \bibnamefont{and}
  \bibinfo{author}{\bibfnamefont{W.}~\bibnamefont{Reynolds}},
  \bibinfo{journal}{J. Comp. Phys.} \textbf{\bibinfo{volume}{{\bf 56}}},
  \bibinfo{pages}{272} (\bibinfo{year}{1984}).

\bibitem[{\citenamefont{Xu et~al.}(2007)\citenamefont{Xu, Dong, Maxey, and
  Karniadakis}}]{xuetal07}
\bibinfo{author}{\bibfnamefont{J.}~\bibnamefont{Xu}},
  \bibinfo{author}{\bibfnamefont{S.}~\bibnamefont{Dong}},
  \bibinfo{author}{\bibfnamefont{M.~R.} \bibnamefont{Maxey}}, \bibnamefont{and}
  \bibinfo{author}{\bibfnamefont{G.~E.} \bibnamefont{Karniadakis}},
  \bibinfo{journal}{J. Fluid Mech.} \textbf{\bibinfo{volume}{582}},
  \bibinfo{pages}{79 } (\bibinfo{year}{2007}).

\bibitem[{\citenamefont{Min et~al.}(2006)\citenamefont{Min, Kang, Speyer, and
  Kim}}]{minetal06}
\bibinfo{author}{\bibfnamefont{T.}~\bibnamefont{Min}},
  \bibinfo{author}{\bibfnamefont{S.}~\bibnamefont{Kang}},
  \bibinfo{author}{\bibfnamefont{J.}~\bibnamefont{Speyer}}, \bibnamefont{and}
  \bibinfo{author}{\bibfnamefont{J.}~\bibnamefont{Kim}}, \bibinfo{journal}{J.
  Fluid Mech.} \textbf{\bibinfo{volume}{558}}, \bibinfo{pages}{309}
  (\bibinfo{year}{2006}).

\bibitem[{\citenamefont{Dean}(1978)}]{dean78}
\bibinfo{author}{\bibfnamefont{R.}~\bibnamefont{Dean}}, \bibinfo{journal}{ASME
  J. Fluids Eng.} \textbf{\bibinfo{volume}{100}}, \bibinfo{pages}{215}
  (\bibinfo{year}{1978}).

\bibitem[{\citenamefont{Moser et~al.}(1999)\citenamefont{Moser, Kim, and
  Mansour}}]{mkmdnsdata}
\bibinfo{author}{\bibfnamefont{R.~D.} \bibnamefont{Moser}},
  \bibinfo{author}{\bibfnamefont{J.}~\bibnamefont{Kim}}, \bibnamefont{and}
  \bibinfo{author}{\bibfnamefont{N.~N.} \bibnamefont{Mansour}},
  \bibinfo{journal}{Phys. Fluids} \textbf{\bibinfo{volume}{11}},
  \bibinfo{pages}{943} (\bibinfo{year}{1999}),
  \urlprefix\url{http://turbulence.ices.utexas.edu/MKM_1999.html}.

\bibitem[{\citenamefont{Iwamoto et~al.}(2002)\citenamefont{Iwamoto, Suzuki, and
  Kasagi}}]{kasagidnsdata}
\bibinfo{author}{\bibfnamefont{K.}~\bibnamefont{Iwamoto}},
  \bibinfo{author}{\bibfnamefont{Y.}~\bibnamefont{Suzuki}}, \bibnamefont{and}
  \bibinfo{author}{\bibfnamefont{N.}~\bibnamefont{Kasagi}},
  \bibinfo{journal}{Int. J. Heat and Fluid Flow} \textbf{\bibinfo{volume}{23}},
  \bibinfo{pages}{678} (\bibinfo{year}{2002}),
  \urlprefix\url{http://www.thtlab.t.u-tokyo.ac.jp/DNS/dns_database.html}.

\bibitem[{\citenamefont{Townsend}(1961)}]{townsend61}
\bibinfo{author}{\bibfnamefont{A.}~\bibnamefont{Townsend}},
  \bibinfo{journal}{J. Fluid Mech.} \textbf{\bibinfo{volume}{11}},
  \bibinfo{pages}{97} (\bibinfo{year}{1961}).

\bibitem[{\citenamefont{Brouwers}(2007)}]{brouwers07}
\bibinfo{author}{\bibfnamefont{J.~J.~H.} \bibnamefont{Brouwers}},
  \bibinfo{journal}{Phys. Fluids} \textbf{\bibinfo{volume}{19}},
  \bibinfo{eid}{101702} (\bibinfo{year}{2007}).

\bibitem[{\citenamefont{Lundgren}(2007)}]{lundgren07}
\bibinfo{author}{\bibfnamefont{T.~S.} \bibnamefont{Lundgren}},
  \bibinfo{journal}{Phys. Fluids} \textbf{\bibinfo{volume}{19}},
  \bibinfo{eid}{055105} (\bibinfo{year}{2007}).

\bibitem[{\citenamefont{Gioia and Chakraborty}(2006)}]{gioiachakraborty06}
\bibinfo{author}{\bibfnamefont{G.}~\bibnamefont{Gioia}} \bibnamefont{and}
  \bibinfo{author}{\bibfnamefont{P.}~\bibnamefont{Chakraborty}},
  \bibinfo{journal}{Phys. Rev. Lett.} \textbf{\bibinfo{volume}{96}},
  \bibinfo{eid}{044502} (\bibinfo{year}{2006}).

\bibitem[{\citenamefont{Mehrafarin and
  Pourtolami}(2008)}]{mehrafarinpourtolami08}
\bibinfo{author}{\bibfnamefont{M.}~\bibnamefont{Mehrafarin}} \bibnamefont{and}
  \bibinfo{author}{\bibfnamefont{N.}~\bibnamefont{Pourtolami}},
  \bibinfo{journal}{Phys. Rev. E} \textbf{\bibinfo{volume}{77}},
  \bibinfo{eid}{055304(R)} (\bibinfo{year}{2008}).

\bibitem[{\citenamefont{Townsend}(1976)}]{townsend76}
\bibinfo{author}{\bibfnamefont{A.}~\bibnamefont{Townsend}},
  \emph{\bibinfo{title}{{The Structure of Turbulent Shear Flow}}}
  (\bibinfo{publisher}{Cambridge University Press}, \bibinfo{year}{1976}).

\bibitem[{\citenamefont{Barenblatt et~al.}(1997)\citenamefont{Barenblatt,
  Chorin, and Prostokishin}}]{barenblattetal97}
\bibinfo{author}{\bibfnamefont{G.}~\bibnamefont{Barenblatt}},
  \bibinfo{author}{\bibfnamefont{A.}~\bibnamefont{Chorin}}, \bibnamefont{and}
  \bibinfo{author}{\bibfnamefont{V.}~\bibnamefont{Prostokishin}},
  \bibinfo{journal}{Proc. Natl. Acad. Sci.} \textbf{\bibinfo{volume}{94}},
  \bibinfo{pages}{773} (\bibinfo{year}{1997}).

\bibitem[{\citenamefont{George}(2007)}]{george07}
\bibinfo{author}{\bibfnamefont{W.~K.} \bibnamefont{George}},
  \bibinfo{journal}{Phil. Trans. R. Soc. A} \textbf{\bibinfo{volume}{365}},
  \bibinfo{pages}{789} (\bibinfo{year}{2007}).

\bibitem[{\citenamefont{Press et~al.}(1996)\citenamefont{Press, Teukolsky,
  Vetterling, and Flannery}}]{nr}
\bibinfo{author}{\bibfnamefont{W.~H.} \bibnamefont{Press}},
  \bibinfo{author}{\bibfnamefont{S.~A.} \bibnamefont{Teukolsky}},
  \bibinfo{author}{\bibfnamefont{W.~T.} \bibnamefont{Vetterling}},
  \bibnamefont{and} \bibinfo{author}{\bibfnamefont{B.~P.}
  \bibnamefont{Flannery}}, \emph{\bibinfo{title}{{Numerical recipes in Fortran
  77}}} (\bibinfo{publisher}{Cambridge University Press},
  \bibinfo{year}{1996}).

\bibitem[{\citenamefont{Schmelcher and Diakonos}(1998)}]{schmelcherdiakonos98}
\bibinfo{author}{\bibfnamefont{P.}~\bibnamefont{Schmelcher}} \bibnamefont{and}
  \bibinfo{author}{\bibfnamefont{F.~K.} \bibnamefont{Diakonos}},
  \bibinfo{journal}{Phys. Rev. E} \textbf{\bibinfo{volume}{57}},
  \bibinfo{pages}{2739} (\bibinfo{year}{1998}).

\end{thebibliography}
\end{document}